\documentclass{article} 
\usepackage{amsmath}
\usepackage{graphicx}
\usepackage{ifthen}
\usepackage{hhline}
\graphicspath{{./figs/}}
\usepackage[backend=biber]{biblatex}
\bibliography{teaching}
\usepackage[margin=2cm]{geometry}

\title{Introducing the Physics of Complex Systems through Videogames}

\newcommand{\orcidA}{$^{0000-0000-0000-000X}$} 
\newcommand{\orcidB}{$^{0000-0002-6293-0305}$} 
\newcommand{\orcidC}{$^{0000-0002-0203-4461}$}
\newcommand{\orcidD}{$^{0000-0003-0543-4900}$}

\author{
\begin{minipage}{0.8\columnwidth}
\begin{center}
Alessio Focardi $^{1}$\orcidA{}, \\
Franco Bagnoli $^{1,4,5,6}$\orcidB{}, \\
Andrea Guazzini $^{2,4}$\orcidC{}\\
Giorgio Gronchi $^{3,4}$\orcidD{}
\end{center}
$^{1}$ \quad Department of Physics and Astronomy, University of Florence, via G. Sansone 1, 50019 Sesto Fiorentino (Italy); 
alessio.focardi@complexworld.net, franco.bagnoli@unifi.it\\
$^{2}$ \quad Department of  
Literatures, Intercultural Studies, Languages and Psychology, University of Florence (Italy); 
andrea.guazzini@unifi.it\\
$^{3}$ \quad Department of
Neuroscience, Psychology, Drug Research and Child Health, University of Florence (Italy); 
giorgio.gronchi@unifi.it\\
$^{4}$ \quad Center for the Study of Complex Dynamics, University of Florence (Italy); \\
$^{5}$ \quad National Institute of Nuclear Physics INFN, Sez. Firenze\\
$^{6}$ \quad UMR ESPACE-Dev, University of Perpignan via Domitia UPVD, 52 Avenue Paul Alduy, Perpignan 66000 (France)
\end{minipage}
}

\begin{document}
\maketitle

\abstract{The purpose of this work is to explore a teaching methodology aimed at communicating topics and subjects not typically studied and analyzed in the (Italian) secondary school. We focused specifically on the use of videogames as a recreational and educational tool, grounding our approach in a broadened conceptual view in which engagement and attentional allocation interact with motivational and affective components of play. Within this perspective, the playful format is considered not only to enhance motivation and enjoyment, but also to attenuate learners' counter-arguing tendencies when confronted with unfamiliar or abstract material. Building on this framework, we developed or adapted several videogames whose central scientific topics are phase transitions, sensitivity to initial conditions, and synchronization. We had a certain number of high school students playing the games, and we asked them several questions to guide them and determine whether the communication was successful. At the end of the activity, we administered a questionnaire about the enjoyment and the difficulties encountered in each game, and the relevant critics. }

\paragraph{keywords:} Gamification, Physics of Complex Systems, Educational videogames, Agent-based simulations

\let\at@
\catcode`@=\active
\def@#1{\ifmmode\boldsymbol{#1}\else\at#1\fi}

\let\quot"
\catcode`"=\active
\def"#1"{``#1''}

\renewcommand\P{\mathcal{P}}
\renewcommand\H{\mathcal{H}}
\newcommand\Z{\mathcal{Z}}

\newcommand{\eq}[2][]{%
    \ifthenelse{\equal{#1}{}}{%
        \begin{equation*}
            #2%
        \end{equation*}%
    }{%
        \begin{equation}\label{eq:#1}%
            #2%
        \end{equation}%
    }%
}
\newcommand{\meq}[2][]{%
    \ifthenelse{\equal{#1}{}}{%
        \begin{equation*}%
            \begin{split}%
                #2%
            \end{split}%
        \end{equation*}%
    }{%
        \begin{equation}\label{eq:#1}%
            \begin{split}%
                #2%
            \end{split}%
        \end{equation}%
    }%
}

\newcommand{\De}[3][]{\frac{\partial^{#1} #2}{\partial #3^{#1}}}
\newcommand{\eps}{\varepsilon}

\newcommand{\Eq}[1]{Eq.~\eqref{eq:#1}}

\section{Introduction}

The introduction of topics related to complex systems in the high school science program is  reputed to be important to understand the world we live in, from  biology to life sciences, Earth sciences, social sciences, economics, and so on~\cite{UnderstandingComplexSystems,SustainabilityComplexSystems}. There have also been efforts to introduce these topics to early students~\cite{K12ComplexSystems}, but often computer skills are needed~\cite{ComplexSystemsSupport}, and  some systematic classifications of what a complex system is can be welcomed. 

Something is reputed to be complex if its reaction to a stimulus cannot be simply forecast. One trivial definition identifies complexity with difficulty: if a system is composed of many parts,  each of which has a complicated behavior, it is not surprising if the resulting response is unpredictable. 

In many cases the opposite happens: the superposition of many complex behaviors, like the trajectories of molecules in a gas, can result in simple collective laws (like then laws of ideal gasses). This is due to averaging, and at the end to the central limit theorem, if the parts are uncorrelated. However, if the parts behave coherently, the results cannot be easily decomposed. 

To put this aspect into evidence, it is important to make use of systems composed by many, simple interacting elements. 

The first item that we shall take into consideration concerns phase transitions: given some parameters that deal with the correlation among parts, one can observe the separation of the system into many incoherent parts, or a clustering of almost all the elements into a large collective and coherent ensemble. At transition, we have strong variations. For large systems, the transition can be quite sudden, denoting a non-linear response.

These are generally stochastic systems, and another interesting aspect is the role of the noise: by examining such examples, students realize that each sample is different, but there are common traits and one must perform averages. 

A second class of complex systems are chaotic ones, in which a small variation in the initial conditions triggers quite a different behavior (sensitivity to initial conditions). However, the study of such systems is quite difficult, since  overall behavior is similar (trajectories always stay in the same attractor), so that one must introduce the distance between trajectories, and other mathematical instruments. We have chosen to explore systems that are not  chaotic (their final state is a fixed point), but whose basins of attraction have a fractal shape, so that the sensitivity with respect to initial conditions is preserved. These systems allow also to discuss  the topic of deterministic dynamics (reproducibility of a trajectory). 

Finally, we explored the field of synchronization of oscillators, again a parameter-dependence that promotes the switch from an incoherent behavior to a collective one. Differently from standard phase transition,here we have coupled deterministic systems, so that these topics can be considered as a liaison between the first two topics.  

As far as methodology is concerned, we have dealt  with the use of videogames as a playful and educational tool. 

In designing our educational intervention, we considered not only the disciplinary relevance of complex-systems topics but also the psychological mechanisms that may support learning through playful environments. Although gamification is often associated with increases in motivation, engagement, and attentional focus~\cite{mee2020role}, the cognitive processes through which games facilitate understanding remain only partially clarified in the literature. Research on attention and cognitive effort suggests that engagement reflects a selective reallocation of limited cognitive resources toward task-relevant information, particularly in interactive and multisensory contexts~\cite{clark2016digital, plass2015foundations}. Videogames may therefore sustain attentional allocation and reduce the cognitive load typically associated with abstract reasoning.

A complementary line of work concerns the motivational dynamics elicited by play. Studies grounded in Self-Determination Theory show that videogames excel in satisfying basic psychological needs such as autonomy, competence, and relatedness~\cite{gee2012glued, przybylski2010motivational}. These mechanisms help explain the persistence and immersion often observed in ludic contexts. However, beyond these motivational aspects, we adopted a broader conceptual view informed by research on counter-arguing and cognitive resistance. Evidence from entertainment-education and narrative engagement shows that enjoyable and absorbing activities can attenuate counter-arguing, reduce cognitive dissonance, and lower reactance when individuals encounter unfamiliar or demanding material~\cite{slater2002entertainment, moyer2008toward, green2000role}. From a cognitive resource perspective, reducing defensive processing frees attentional capacity that can be devoted to assimilating the conceptual structure of the task~\cite{petty1986elaboration, wegener1995positive}.

An additional theoretical element guiding our approach concerns cognitive ergonomics. In this perspective, videogames function not merely as motivational amplifiers, but as environments that structure perception, action, and reasoning through implicit rules, feedback loops, and goal hierarchies. Such features can reduce the cognitive cost of problem access, providing learners with an intuitive scaffolding that mirrors core logical or dynamical principles inherent to the phenomena being modeled~\cite{norman1988psychology, van2024ten}. By embedding the study of complex systems within game mechanics that naturally encode interaction rules, state transitions, and cause–effect relations, the activity may support comprehension through a form of proceduralized conceptual guidance.

Taken together, these strands of research converge on a dual hypothesis: (i) playful interaction enhances attentional engagement and motivational readiness, and (ii) affective immersion and cognitive ergonomics reduce counter-arguing and facilitate access to abstract or nonlinear concepts. Before developing the videogames described in this work, we explicitly adopted this expanded psychological framework as a guiding principle, aiming to design environments that could both sustain attention and ease the conceptual entry into the physics of complex systems.

In fact,  videogames have many aspects that can make them a functional didactic learning method since they are multi-sensory (visual, tactile, verbal, and graphic). The playful nature of the game also makes it possible for the class to be actively involved, lightening learning~\cite{GamificationTeaching}.

Both the aspects of the use of (video) games for teaching, and the teaching of physics of complex systems are not new, see for example the activity of the Interdepartmental Research Center Games for Social Change (GiX) of the University of Florence~\cite{GiX}, or the recent thesis work of Veronica Ilari~\cite{Ilari2024}.

To achieve this goal, we have developed or adapted videogames whose central scientific topic is not included in the secondary school curriculum, such as nonlinear phenomena, sensitivity to initial conditions, and phase transitions. The design of the educational proposal then followed. 

A sample of students from all the five-year classes of the science and the social-economics curricula of a Florentine school participated in the game. We let them play with the games for about two hours. During this activity, some questionnaires have been administered in order to guide the players and to check if the games were understood. 

Although the main goal of this experiment was to trigger curiosity and discussions about the relevant aspects and how they are related to other disciplines, at the end of the activity we then sent a questionnaire concerning questions among which those relating to the liking and difficulty found in each individual game stand out, also allowing the possibility of making criticisms. From the analysis of their answers, we are happy to be able to say that all the games proposed were liked, despite the obstacles that students say they found in their development. 

From the criticisms, both positive and negative, made by students, it emerges both the desire to carry out school activities other than frontal teaching, and the pleasure in having fun while learning. One of the biggest difficulties was to concentrate the activity in less than two hours and, probably, a structured activity of this type could take twice as long to be carried out in the most optimal way possible. 

A longer interaction time could provide students with the opportunity to become more familiar with the dynamics of the game, also creating a stronger bond between teacher and class. In the future, therefore, we plan to continue to offer the activity to other schools as well, requesting longer time windows or dividing the activity into several game sessions, and focusing more on the educational (for example, leaving more time for discussion and peer instruction) and playful aspects, reducing the questions asked in writing.

In this work we will first introduce some physical models on which some videogames we have developed are based, together with a description of the videogames themselves and the educational proposal we have created. We will then deal with the experimental procedure that we have implemented in our research: presentation of videogames in high school classes of scientific and social economic address, questions asked to students anonymously and finally an analysis of the answers received.

The scheme of this paper is the following. In Sect.~\ref{sec:fire}, we describe the first set of games, related to phase transitions and in which a forest fire spreading is modeled. The second set of games, dedicated to the simulation of the chaotic pendulum is illustrated in Sect.~\ref{sec:pendulum}. In Sect.~\ref{sec:fireflies}, the Kuramoto systems, modeled as a firefly synchronization problem, is presented. 

We made use of the NetLogo platform~\cite{NetLogo}, which has several advantages with respect to developing custom games, as illustrated in Sect.~\ref{sec:NetLogo}. A similar approach has also been followed by others~\cite{starlogo}. 

NetLogo is interesting as an educational tool because it not only allows you to easily develop an interactive interface but also provides the source code and the ability to modify it. It is also possible to ``export'' the platform in html + javascript format, so that students can also use the platform from mobile phones or tablets, although the capabilities of this version are still quite limited. We added the possibility of recording data through a Google form. See Sect.~\ref{sec:NetLogo} for the implementation details.

Finally, in Sect.~\ref{sec:results} we report about the data and the feedback collected. Conclusions are drawn in the last section.

This work is based on the master thesis in Physics by A. Focardi~\cite{Focardi2024}.

\section{Forest fire}\label{sec:fire}

The study of the spread of a fire in a forest or in a woodland can be very interesting both from a mathematical point of view and from a more concrete point of view, as it can help to understand which reforestation techniques and strategies to apply in order to prevent a fire that spans the entire forest~\cite{albinet86,beer1990fire,Drossel1992, galeano2015}.

\section{Forest fire: the physics of the problem}\label{sec:fire-physics}

In the context of physics and mathematics, the forest fire spread is described as a percolation model, a theory that refers to the behavior of some \textit{clusters} in a given lattice whose sites are occupied or bonds are present with a certain probability~\cite{kirkpatrick1960,Essam1980,Stauffer2018}.

It is interesting to use percolation models since there are three possible fire propagation regimes.
\begin{itemize}
    \item No spread. In this situation, the trees are so separated from each other, or the transmission probability is so small that the fire can only burn a small portion of the forest. This implies that the fire starting from a different initial point will affect a different group of trees, therefore the forest is divided into uncorrelated clusters. One obtains different but qualitative similar results after repeating the simulation. 
    \item Widespread fire. In this case, the fire doesn't stop and manages to burn almost the entire forest. This implies that in this case a large part of the forest is correlated. Again, results do not differ much in repetitions.
    \item Critical region, for intermediate values of the parameters. In this case, we have clusters of all sizes, with very complex (fractal) shapes. Repeating the experiment can give very different results and also measured quantities (like the percentage of burned trees) show large fluctuations.
\end{itemize}
    
Percolation models are important because they are able to describe many different phenomena: we have introduced them as a model of forest fire, but they can also serve as model of infiltration of a liquid in a porous material, diffusion of an illness in a cellular matrix, diffusion of opinions or news in a contact networks, and so on. 

If viewed as an epidemic process, our model corresponds to a SIR (susceptible-infected-removed) model~\cite{Harko2014}. 
    
In our context, the terrain is divided into a square lattice of cells, each of which can be in four states: empty, occupied by a healthy tree, occupied by a burning tree, and occupied by an already burned tree, which, in terms of propagation, behaves like an empty cell.
    
The simplest rule is that a burning tree reliably communicates its burning to all neighboring cells occupied by a healthy tree, but we also considered the case in which this propagation occurs probabilistically, justified saying the smaller trees are not always in touch with neighboring ones.
This second parameter is related to the presence or absence of bonds between cells.

The fire is linked to a site percolation problem because in this model, a fire can spread to a substantial fraction of the forest only if there is a cluster of healthy trees spanning a significant fraction of the grid (the "giant component" of percolation theory).

At beginning, the grid contains only empty or tree-occupied cells, and a random cell is set on fire.
    
If the probability $p_s$ that a site is occupied by a tree is large enough, a ``giant'' cluster will exist and the initial fire site will most likely be part of it. Conversely, if this probability is low, only small, disconnected clusters will exist.
    
\begin{figure}[ht]
    \centering   \includegraphics[width=0.6\textwidth]{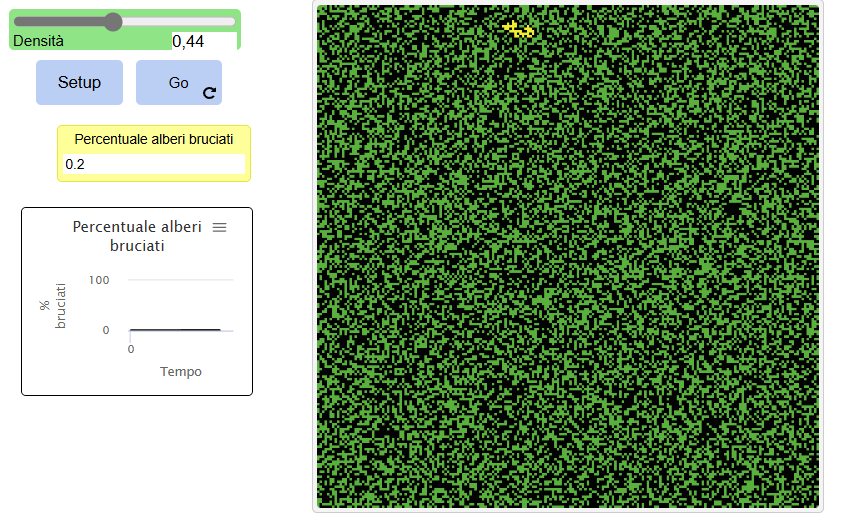}
    \caption{The forest fire game with a density of trees smaller than the critical value. The initial fire cannot spread. }
    \label{fig:FFbassadensita}
\end{figure}

\begin{figure}[ht]
    \centering   \includegraphics[width=0.6\textwidth]{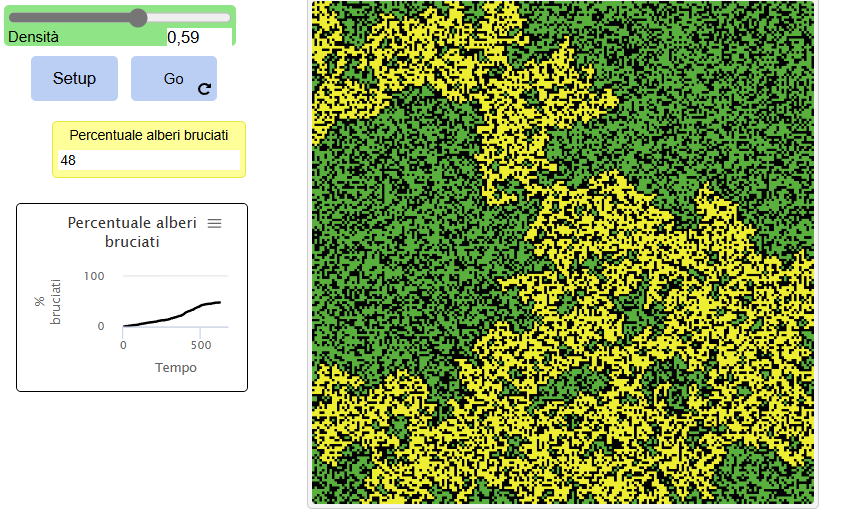}
    \caption{The forest fire game with a critical density of trees. The cluster of burnt trees has a fractal character}
    \label{fig:FFdensitacritica}
\end{figure}

\begin{figure}[ht]
    \centering   \includegraphics[width=0.6\textwidth]{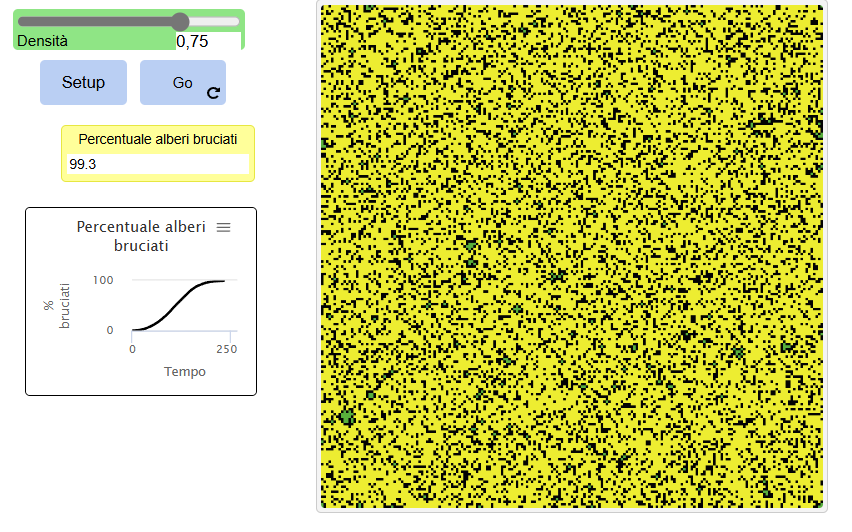}
    \caption{The forest fire game with a density of trees larger than the critical value. Almost all trees burned.}
    \label{fig:FFaltadensita}
\end{figure}

This leads to a very interesting consequence: there is a critical occupancy probability, $p_c$, that divides these two extreme cases (see Figs.~\ref{fig:FFbassadensita}, \ref{fig:FFdensitacritica}, \ref{fig:FFaltadensita}).

A less trivial result is the statistical variability of all length scales and self-similarity. These characteristics highlight the relationship between percolation models and phase transitions.

Our games are based on a two-dimensional square lattices of a certain dimension $L\times L$ whose sites are occupied by trees with a certain probability $p_s$. Since the results in percolation theory depend on both the geometry of the lattice and its dimension, our conclusions make no claims of generality.
    
In our model we only consider interactions between the four closest neighbors (North, South, East, and West), thus not considering sites on the diagonal as neighbors. Furthermore, we will not include some phenomena such as the presence of wind.
    
The possible states are as follows.
\begin{itemize}
    \item Healthy tree (state 0). Not burning but can be set on fire if fire gets close.
    \item ``Disease'' tree (state 1). In this state, the tree is burning and can spread  fire to nearby trees. When there are no more trees in this state, the fire can no longer spread.
    \item Burnt Tree (State 2). Once the tree is no longer in its previous state, it automatically transitions to the burned tree's state, where it cannot re-burn or spread the fire.
    \item Absence of trees (state 3), equivalent to state 2.
\end{itemize}
    
The system setup is such that there is a probability $p_s$ of having healthy trees and $1 - p_s$ of having empty spaces (state 3). At this point, one of the trees in state 0 is randomly chosen and switches to state 1, thus starting the fire.

According to our model, a burning tree can set fire to its nearest neighbors with a certain probability. As already mentioned, what we observe is the existence of a critical probability $p_c$ that divides two extreme cases: the case in which the fire almost doesn't spread at all, and the case in which practically the entire forest burns.
    
Our work confirms the result in the literature~\cite{beer1990fire} for two-dimensional square lattices, with a value for the critical probability equal to $p_c = 0.59$.

We can then complicate the model by adding the probability $p_b$ that a burning tree (state 1) will spread the fire to one of its four possible neighbors.
    
This type of problem is certainly more difficult to deal with than the previous one, since there are two parameters: the probability of occupation of a site ($p_s$), and the probability of propagation ($p_b$).

\begin{figure}[ht]
    \centering   \includegraphics[width=0.6\textwidth]{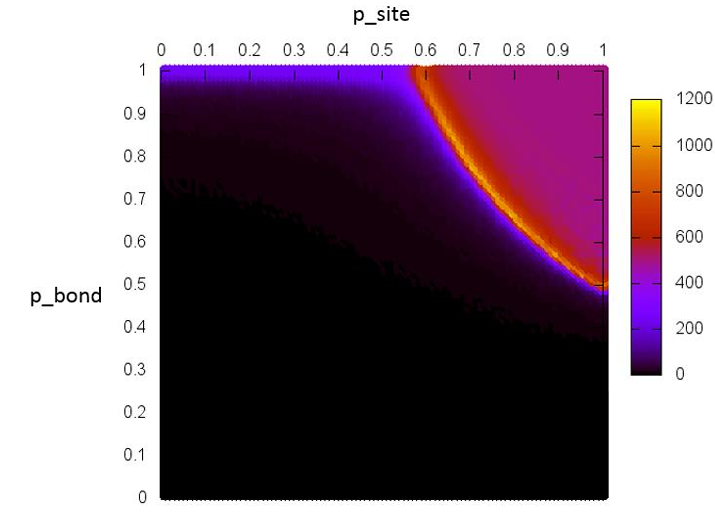}
    \caption{Forest fire phase diagram as a function of tree density (site probability) and dimension (bond probability). }
    \label{fig:Phase}
\end{figure}

If we want to have an immediate idea of the threshold value, we can construct the phase diagram in which both  parameters are present, shown in Fig.\ref{fig:Phase}~\cite{galeano2015}.

\subsection{Forest fire: educational proposal}\label{sec:fire-eucational}

The goal of the game is to demonstrate the phenomenon of phase transitions and the probabilistic nature of certain physical systems. Specifically, this game simulates the spread of a forest fire, which depends only on the numerical density of trees (i.e., the number of trees present on the playground).
    
The starting point was the model \textit{Fire} on the NetLogo platform~\cite{NetLogoFire}.
    
At the beginning of the game, the player chooses the tree density by means of a slider, and the square lattice is randomly populated.
    
One of the trees is then put on fire. The fire can spread only to nearby trees, and the percentage of  burned trees is shown on a plot. We choose zero boundary conditions, which are considered more "natural" (the fire cannot "jump" over opposite sides).
    
The goal is to understand the relationship between the tree density and the percentage of trees burned, their "universal" characteristics (for instance the shape of the plotted curve) and  the "particular" ones (the shape of the burned cluster).

Specifically, the class was asked to infer the tree density that would (approximately) allow half of the forest to be burned. More specifically, they were told that a range between 40\% and 60\% would be acceptable. This must be achieved through various simulations in which the tree density could be adjusted from time to time until the objective was reached. 
    
While the numerical density of trees can be varied via a slider, the percentage of burned trees is  displayed both qualitatively, by observing the game playground, and quantitatively, via a dialog that displays these data and/or via a function graph that illustrates this quantity as it varies over time.

The only adjustable parameter in this game is the numerical density of the trees.
    
The asymptotic percentage of burned trees exhibits a phase transition: below a certain critical density, only a small percentage of trees are finally burned, while if the chosen density is larger than the critical value, this quantity approaches the unity. 

It is interesting to discuss with the class the plot of the percentage of trees burned as a function of time, and it is possible to do so on several levels based on the mathematical preparation of the class itself, moving from a more qualitative explanation to a more accurate approximation of this behavior.

\begin{figure}[ht]
    \centering
    \includegraphics[width=0.5\textwidth]{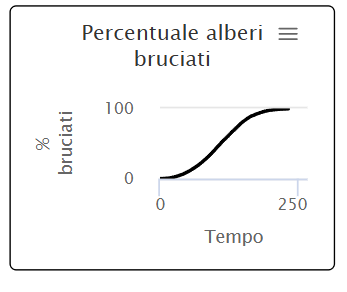}
    \caption{The characteristic S-curve of the percentage of  burned trees as a function of time.}
    \label{fig:CurvaS}
\end{figure}

It is possible to note that for density values smaller than the critical density (see fig.~\ref{fig:FFbassadensita}) the graph is flattened near zero (practically no trees burns), while when the critical density is reached (fig.~\ref{fig:FFdensitacritica}) or exceeded, the typical "S" graph appears in which there is a sharp increase and then a saturation, as shown in fig.~\ref{fig:CurvaS}. 

When the density is quite large, as is the case in fig.~\ref{fig:FFaltadensita}, it may be interesting to highlight the progression of the fire on the map: the diamond-shaped front is due precisely to the particular way in which the fire spreads to its closest neighbors. One could also experiment with different lattice sizes. 
    
This aspect is not the only interesting one, both from a physical and educational point of view. The quantity of burned trees also depends on the position of the tree that starts the fire. For example, if it is close to the edge of the map (due to zero boundary conditions) or is in an area with many ``empty spaces'' around it, the fire does not spread easily.
    
Students are visibly surprised when they observe this type of dynamic. They expect the game (and perhaps even all physics and the laws of nature) to be completely deterministic, and when they see that running multiple simulations using the same density results in different percentages of burned trees, they are perplexed.
    
One aspect on which this \textit{serious game} focuses is also offered by the statistical variability that results from running the simulations multiple times, which is also more marked when the density is close to the critical one.

We also built a variant of the previous game in which the spread of a the fire assumed to depends on two parameters: the numerical density of trees, that is, the number of trees present on the playground, and their size, which reflects the probability of igniting nearby trees.

At the beginning of the game, the lattice is populated by  randomly arranged trees  with a certain numerical density and size that can be controlled by the player.
    
Then one of the trees is randomly chosen and set to fire.   The fire will be able to spread only to nearby trees with a probability that depends on the size of the tree itself: the larger the tree, the higher the probability that the fire will propagate.
    
The goal of the game is to understand the relationship between tree density and size so that the fire only affects half the forest: if the tree density is high, for example, to achieve approximately 50\% burn rate, a relatively low tree size should be chosen (i.e., less likely to set fire to nearby trees). This can be achieved through various simulations in which the physical parameters are varied until the objective is achieved.

The class  is asked to discover some combination of parameters (typically four or five combinations) that allow the burning of (approximately) half of the forest, more precisely it was told that a range between 40$\%$ and 60$\%$ could be acceptable, as in the previous game.    
    
\begin{figure}[ht]
    \centering
    \includegraphics[width=0.5\textwidth]{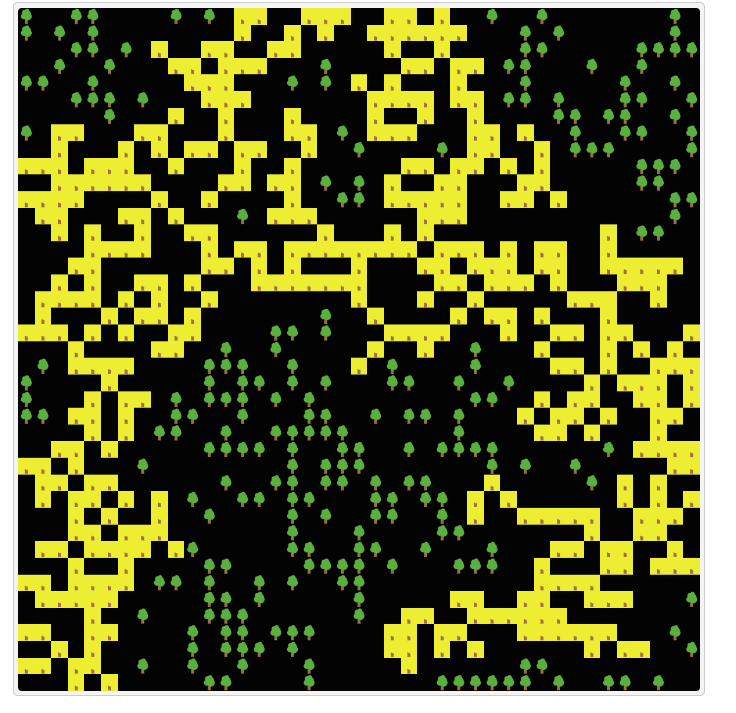}
    \caption{Fire in a forest a tree density of 40\% and a tree size of 92\%.}
    \label{fig:FFP}
\end{figure}

Both the numerical density of trees and their size can be varied using a slider. The percentage of trees that are burned can be read/displayed both qualitatively, by observing the game map (as in fig.~\ref{fig:FFP}), and quantitatively, through a window that displays this data. 

Compared to the previous game, the  additional difficulty lies in the presence of the additional probabilistic parameter (the tree size). This aspect is educationally interesting because it allows us to emphasize the incomplete determination of certain phenomena, and discussion with the class regarding other possible uses of this physical model, such as in epidemiology, is very useful.
    
Also not to be overlooked is the idea that some systems, even very complex or complicated ones, can still be described, within a certain approximation, with very simple models and with few parameters, but which contain the essential ingredients needed to describe ``reality''.
    
In addition to proposing a discussion on the similarities with epidemiological models, a discussion on what other parameters might influence the probability of fire can start. This allows us to demonstrate how the problem can  gradually become more complex.

To conclude this first phase of forest-themed games, a test game was prepared to verify the understanding, even if partial, of the first two games.
    
The playground of this game, also square, is composed of trees with a double symmetric variation: The trees become smaller in the horizontal dimension and sparser in the vertical one, both starting from the center.

\begin{figure}[ht]
    \centering
    \begin{tabular}{cc}
    (a) & (b) \\
    \includegraphics[width=0.4\textwidth]{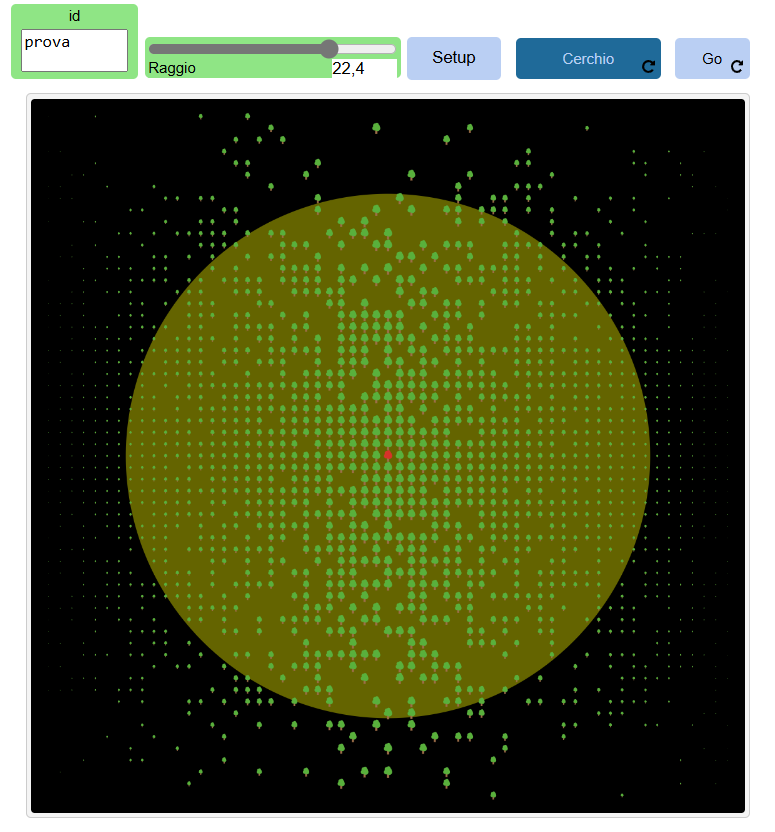} & 
    \includegraphics[width=0.4\textwidth]{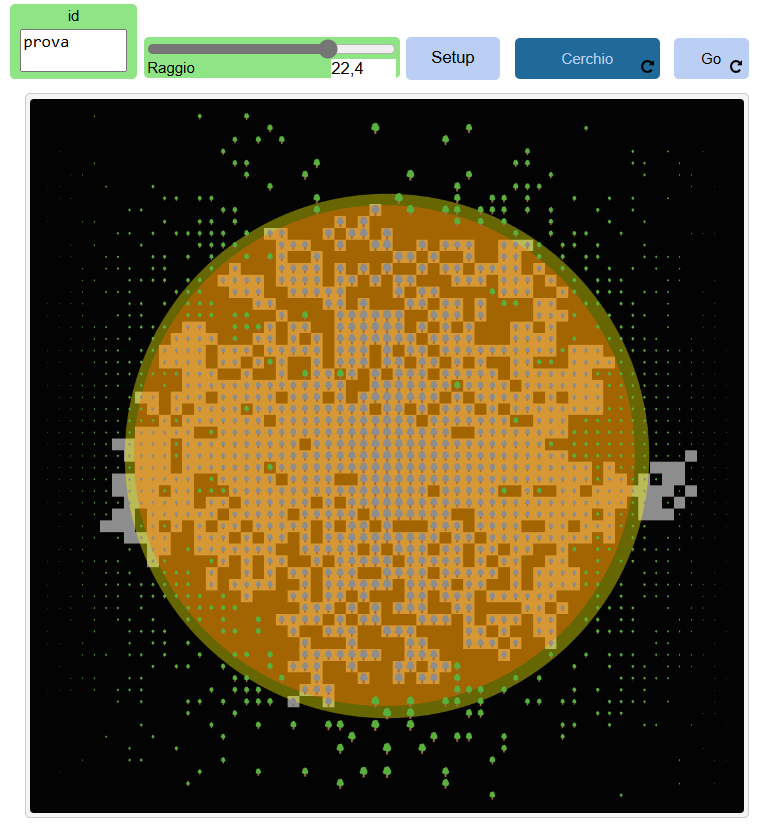}
    \end{tabular}
    \caption{The forest fire test screenshot, (a) with the target area chosen (green circle); (b)  with the target area chosen (green circle), the burned area (white squares) and the "optimal" circle area (reddish circle).}
    \label{fig:FFtest}
\end{figure}

This type of map ensures that the fire only spreads  to a certain average distance from the center, and the class is asked to estimate the  radius of the burned area. By moving a cursor, it is possible to increase or decrease the estimated radius, represented graphically on the map by a circle originating from the central tree, as in Fig.~\ref{fig:FFtest}-(a).

The "optimal" circle has been estimated to be proportional to the gyration radius of the burned area. Indicating by $(x_i, y_i$) the coordinates of the burned trees, by $n$ the number of burned trees and by $(\bar{x}, \bar{y}$) the coordinates of the center of mass,
\meq{
 \bar{x} &= \frac{1}{n} \sum_i x_i,\\
 \bar{y} &= \frac{1}{n} \sum_i y_i,\\
}
the gyration radius $r$ is 
\eq{
    r = \sqrt{\frac{a}{n}  \sum_i \left[(x_i-\bar{x})^2 +(y_i-\bar{y})^2\right] },
}
where the parameter $a$ has been heuristically chosen equal to 2, after testing with 10 volunteers. 

Once the radius has been chosen, the student starts the fire and observes whether their intuition was correct; see, for example, Fig.~\ref{fig:FFtest}-(b). Furthermore, as an output, the game informs the user about the average distance from the center, thus allowing both a graphical self-assessment via the map and a quantitative one.

\subsection{In-depth considerations for the forest-fire problem}\label{sec:fire-depth}

It is possible to derive a simple mean-field description of the phenomena, assuming that there are no spatial correlations, analogous to the standard mathematical treatment of epidemics models~\cite{Kermack1991,Kermack1991a,Kermack1991b}. 

We denote by $S(t)$ the fraction of susceptible (healthy) trees, with $I(i)$ the infected agents (burning trees) and by $R(t)$ the removed trees, including also the empty areas. 

The basic assumption is that the probability of having one of the agent types in a each cell is given by its fraction in the total population. This is equivalent to reshuffling, at each time step, all agents on the lattice, or, equivalently, to considering a well-stirred mixture. In essence, these are the equations used in modeling chemical reactions. 

Susceptible trees convert into infected if there is at least one infected neighbor among the four. Let us denote this probability as $P(S\rightarrow I)$. It is easier to derive the probability of not being infected, $1-P(S\rightarrow I)$, that is given by the probability for a healthy site of not having any infected neighbors. Since we assume that all events are independent, probabilities factorize and therefore, for 4 neighbors, 
\eq{
    1-P(S\rightarrow I) = (1-I)^4.
}
The basic equations (Markov chain) are therefore
\meq[SIR]{
  S(t+1) &= S(t) + S(t)P(S\rightarrow I) =S(t) (1-I(t))^4\\
  I(t+1) &= S(t)P(S\rightarrow I) = S(t) \left(1 - (1-I(t))^4\right)\\
  R(t+1) &= R(t) + I(t)
}
since in our model the infection is certain, and infected agents convert into removed ones in one step. 

\begin{figure}
    \centering
    \begin{tabular}{cc}
    (a) & (b) \\
    \includegraphics[width=0.4\linewidth]{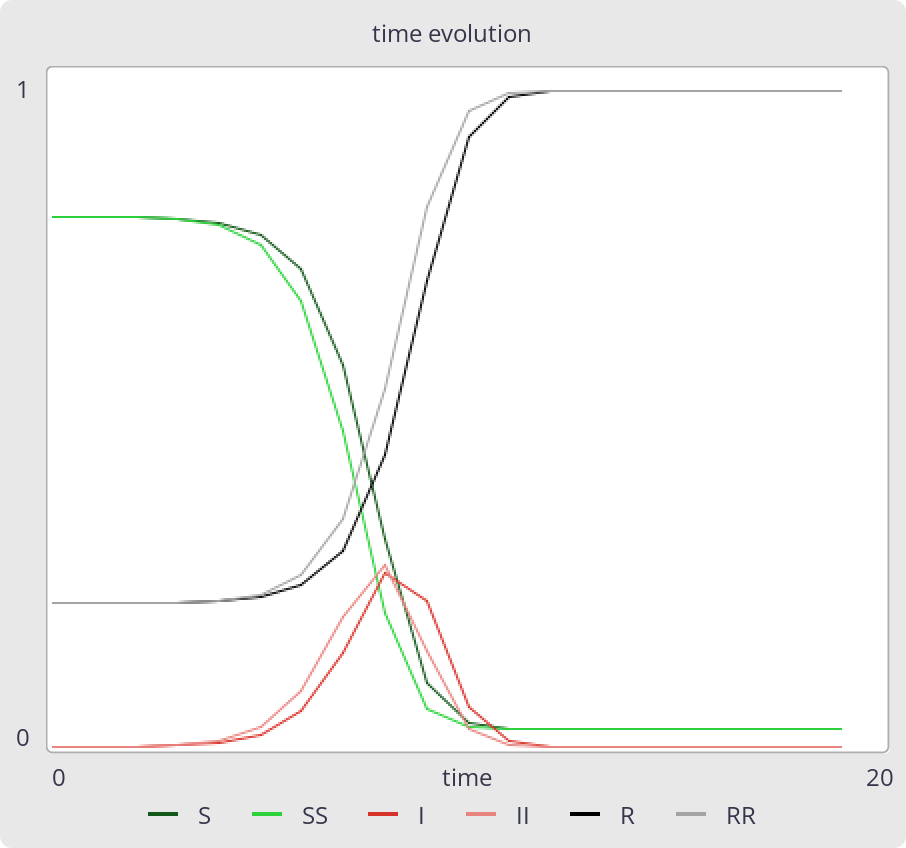} &
        \includegraphics[width=0.4\linewidth]{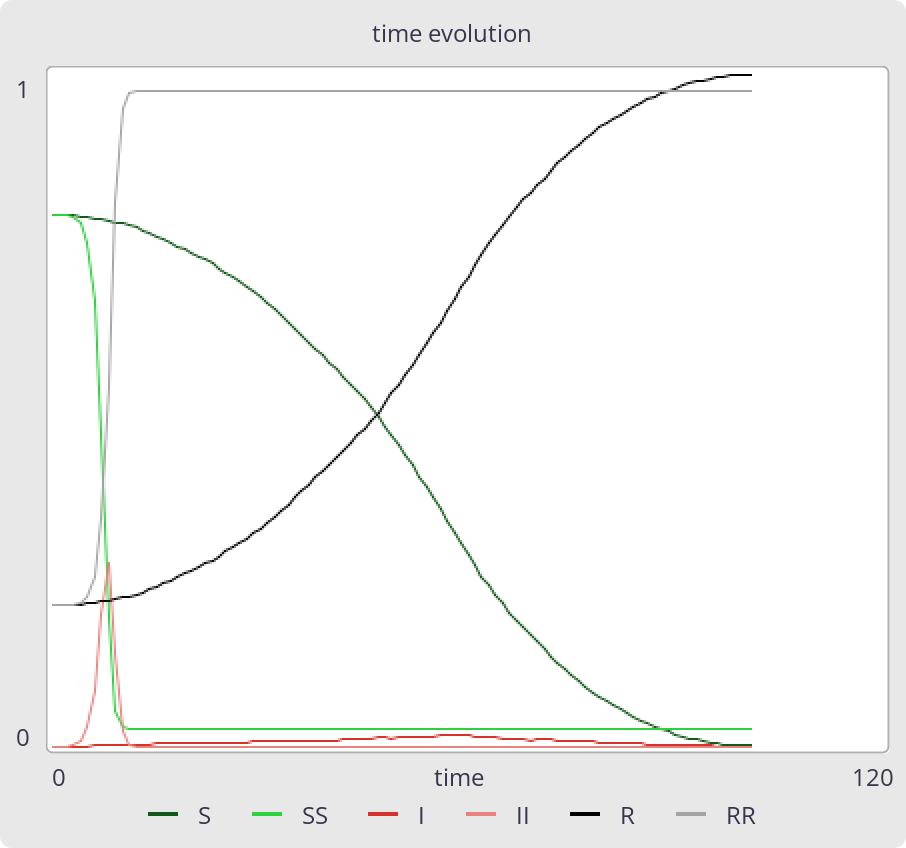}
    \end{tabular}
    \caption{SIR-Forest Fire model. $S,I,R$ denote the quantities measured,  $SS,II,RR$ are the numerical integration of Eq{SIR}. (a) Shuffling the position of trees at each time step. (b) Without shuffling (actual model).}
    \label{fig:SIR}
\end{figure}

One can easily check that the total population $S+I+R$ remains constant in time. 
We can then set up a simulation (SIR) in which cells are shuffled or not at each time step (Fig.~\ref{fig:SIR}). One can notice that this shuffling makes the mean-field approximation to reproduce well the experimental results, while they are quite different without shuffling, marking the importance of the spatial aspects in such models.  

Another interesting experiment is about the finite-size effects. Equation~\eqref{eq:SIR} is derived assuming that the size of the system is infinite, and the resulting critical value of the initial density is zero, since the basic reproduction number (the number of neighbors) is 4 (greater than one). Even with shuffling, the numerical simulations reveals the existence of an effective critical value, that decreases with the system size.

\section{Chaotic magnetic pendulum}\label{sec:pendulum}

\begin{figure}[ht]
    \centering
    \includegraphics[width=0.3\textwidth]{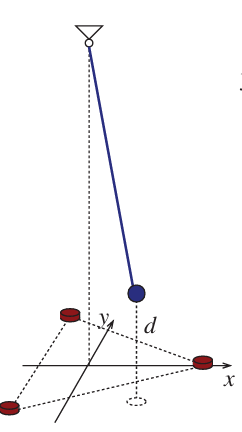}
    \caption{The chaotic magnetic pendulum.}
    \label{fig:pendolo}
\end{figure}

Let us now analyze a system formed by a strong magnet attached to a string,  which  oscillates above a plane on which three other magnets, arranged at the vertices of an equilateral triangle, lie, as shown in Fig.~\ref{fig:pendolo}~\cite{Motter2013}.

The pendulum moves under the influence of gravity, the tension in the string, the resistance due to air friction, and the attractive force due to magnets.

This type of system is very simple from a mechanical point of view since it has a low dimensionality and includes rather simple physical laws, at least with the approximations that will be made. It can also be easily constructed, as commented in Sect.~\ref{sec:pendulum-depth}.

\subsection{Chaotic pendulum: the physics of the problem}\label{sec:pendulum-physics}

Despite its simplicity, this system has complex characteristics. The attractors are only fixed points, corresponding to the three magnets, but their basin of attraction is generally fractal, and therefore it is difficult to estimate its shape.

As already mentioned, to describe the system we make some approximations: the magnets are considered point-like and we assume that the length of the pendulum string is much greater than the distance between the magnets, which allows us to use the approximation of small angles, and to approximate the fact the end of a pendulum lies on a sphere by the tangent plane.

As regards the resistance due to air friction, assuming that the body moves at low speeds (Reynolds number less than 1, i.e. Stokes regime), we can state that the force is proportional to the speed and therefore use, in the case of a spherical mass, Stokes' law,
\[
\vec{F} = - 6 \pi \mu R \vec{v}.
\]

Finally, adding the forces between the magnets and the pendulum, which we assume depends on the inverse square of the distance, we can write the following equations of motion:
\meq{
    \ddot{x} &= - \omega_0 x^2 - \alpha \dot{x} + K \sum_{i = 1}^3 \frac{(\tilde{x}_i - x)}{D_i(\tilde{x}_i, \tilde{y}_i)^3},\\
    \ddot{y} &= - \omega_0 y^2 - \alpha \dot{y} + K \sum_{i = 1}^3 \frac{(\tilde{y}_i - y)}{D_i(\tilde{x}_i,\tilde{y}_i)^3},
}
where $\omega_0^2 = \frac{g}{l}$ is the natural frequency of the pendulum, $\alpha$ is the friction coefficient. $D_i (\tilde{x_i}, \tilde{y_i})$ is the distance between the pendulum and the $i$-th magnet,
\eq{
    D_i (\tilde{x_i}, \tilde{y_i})= \sqrt{(\tilde{x}_i - x)^2 + (\tilde{y}_i - y)^2 + z^2},
}
where $z$ it is the distance between the pendulum and the plane (actually a spherical surface) where the magnets lie.

We choose the reference system so that the coordinates of the magnets on the plane are given by
\eq{
    (\tilde{x}_1, \tilde{y}_1) = \left(\frac{1}{\sqrt3}, 0 \right), (\tilde{x}_2, \tilde{y}_2) = \left(-\frac{1}{2 \sqrt3}, - \frac{1}{2} \right), (\tilde{x}_3, \tilde{y}_3) = \left(-\frac{1}{2 \sqrt3}, \frac{1}{2} \right).
}

With this choice, the system has four fixed points, one unstable (the origin of the reference frame) and three stables, and therefore has three attractors. Since the system is dissipative, all trajectories converge to a fixed point (except for some initial conditions with zero measure).

\begin{figure}[ht]
    \centering
    \begin{tabular}{cc}
    (a) & (b) \\
    \includegraphics[width=0.3\textwidth]{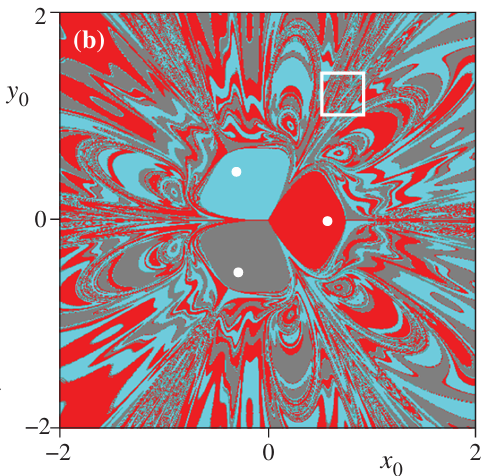} & 
    \includegraphics[width=0.65\textwidth]{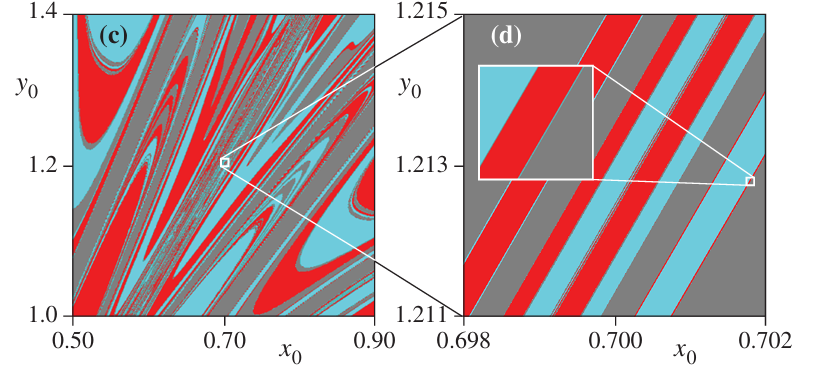}
    \end{tabular}
    \caption{(a) The basins of attraction of the chaotic magnetic pendulum. (b) Zooming in reveals a fractal structure. }
    \label{fig:bacino}
\end{figure}

Figure~\ref{fig:bacino} shows the basins of attraction of the three fixed points considering trajectories with zero initial velocity~\cite{Motter2013}. The colors indicate initial conditions associated with the three different attractors; further magnification reveals structures that blend at increasingly smaller scales, suggesting a fractal structure and a strong sensitivity to the initial conditions.

For further information on the dynamics of this system, we recommend reading Refs.~\cite{Motter2013,zhang2023}.

\subsection{Chaotic pendulum: educational proposal}\label{sec:pendulum-educational}

The aim of the game is to demonstrate the strong dependence on initial conditions present in some physical systems.

The playground is formed by a square lattice (initially all cells are black) on which the three magnets lie, marked by disks of different colors (red, blue and green), while the pendulum is represented by a white disk.

Students are asked to play freely, letting the pendulum start from different positions. When the pendulum reaches one of the three magnets, the square corresponding to the pendulum's starting position is colored with the same color as the target magnet.

\begin{figure}[ht]
    \centering
    \begin{tabular}{cc}
    (a) & (b) \\
    \includegraphics[width=0.4\textwidth]{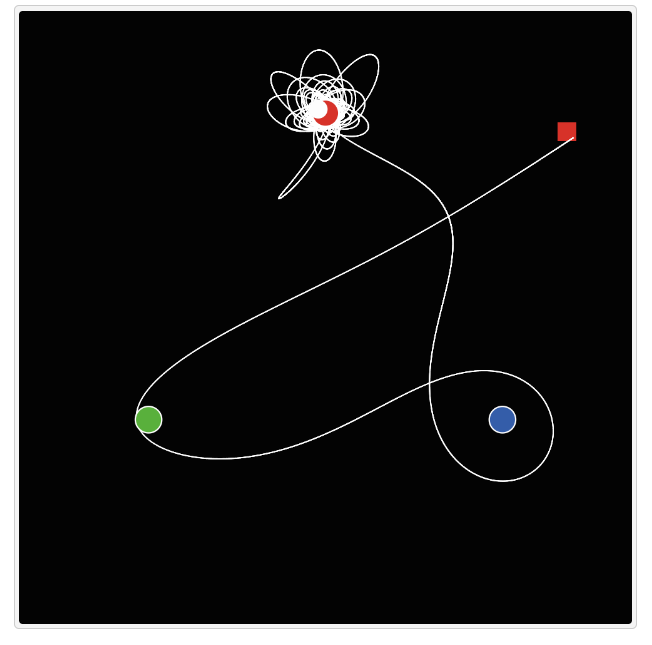} & 
        \includegraphics[width=0.4\textwidth]{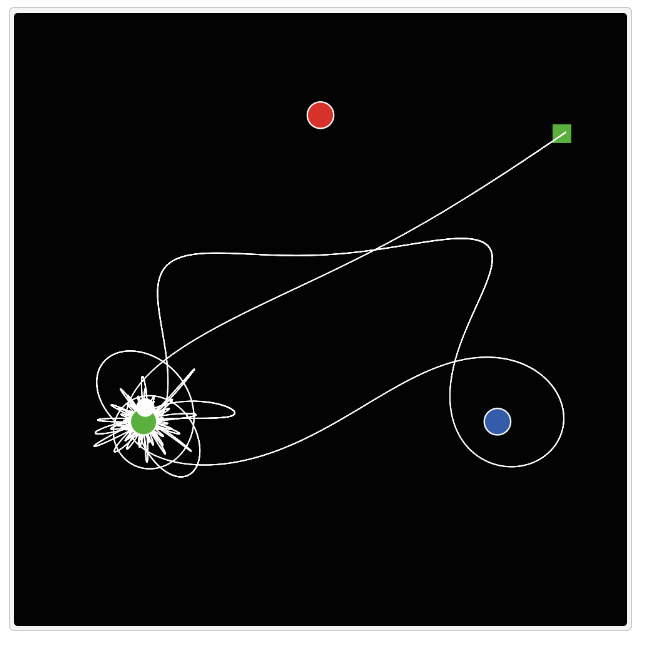}
    \end{tabular}
    \caption{The chaotic pendulum playground, with the colored circles representing magnets and the white circle representing the pendulum. (a) The solid line is the pendulum's trajectory, and the red square provides information about the pendulum's starting and ending points. (b) The trajectory obtained by slightly changing the initial position. }
    \label{fig:CP}
\end{figure}

After a few trials, students notice some peculiarities: there are starting areas in which the pendulum follows very similar trajectories and always ends up on the same magnet, and there are starting areas in which the pendulum follows very unusual curves, like the one in Fig.~\ref{fig:CP}-(a).

Notice that, since the squares have finite width, it may happen that starting from the same square the pendulum ends on a different magnet, thus switching the color of the square. 

To help understand the game better, some clues can be given in the form of questions such as \textit{What happens if the pendulum starts from exactly the same position?} \textit{What happens if it starts from a point very close to the one it started from previously?}

In this way, students can understand that in the first case, since the initial conditions are identical, the trajectories are always the same (determinism), while in the second case the trajectories can change significantly for some selected points (sensitivity to the initial conditions), as can be seen in Fig.~\ref{fig:CP}-(b).

As a class, one can then reflect on how these types of dynamics are almost never covered in curriculum, and discuss other physical examples, including those related to pop culture, such as the three-body problem sci-fi book~\cite{Liu2014-vc}.

The final test to check if they understood the relevant topics is the following:
using the same  playground, students are asked to choose a point on the map that they consider particularly sensitive to the initial conditions, justifying their response.

\subsection{In-depth considerations for the chaotic magnetic pendulum}\label{sec:pendulum-depth}
 
It is not difficult to construct a physical realization of the system, see for instance Ref.~\cite{chaoticpendulum}. One can use the cover of a printer paper box, gluing the magnet to the bottom, so that the top is free. A piece of paper can then be fixed on it, marking the position of the magnets with spots of different colors. Students can then reproduce the simulations, marking with a pencil the initial position of the pendulum, letting it go and finally, using a color marker, assigning to the starting point the color of the  magnet corresponding to the final position.

\section{Fireflies synchronization}\label{sec:fireflies}

The goal of this model is to show how transitions like phase transition can also arise in deterministic systems. 

For this section we will mainly refer to Strogatz's article ``From Kuramoto to Crawford: exploring the onset of synchronization in populations of coupled oscillators''~\cite{Strogatz2000} in which a historical and physical reconstruction of the research on the Kuramoto model is carried out.

The Kuramoto model is composed by many coupled  oscillators whose natural frequencies are given by a certain distribution. These oscillators are coupled to the average value of their phase (a mean-field or all-to-all coupling). 

If the coupling parameter exceeds a certain critical threshold, the system undergoes a phase transition whereby the oscillators synchronize at a certain frequency.

The examples of the use of this model are many and varied, ranging from biology (the synchronization of the light signal of fireflies that we discuss in the game, but also the unison chirping of crickets), medicine (synchronization of heart cells), physics and engineering (from lasers to microwaves to superconducting Josephson junctions).

\subsection{Fireflies synchronization: Physics of the Problem}\label{sec:fireflies-physics}

The phenomenon of collective synchronization was first studied by Wiener~\cite{wiener1961,Wiener66}, but a more sophisticated mathematically approach was provided by Winfree~\cite{winfree67}, who assumed that the oscillators were nearly identical and that their coupling was weak. On small time scales, the oscillators relax to their limit cycles and can be described by considering only their phases; conversely, on long time scales, these phases evolve due to the interaction between the weak coupling and the frequency differences between the oscillators. A further simplification is the one analogous to a mean field in physics, i.e., to consider oscillators coupled all-to-all.
  
Through numerical simulations and some analytical approximations, Winfree discovered that the system can exhibit a kind of phase transition in which, for small coupling, the system behaves incoherently and each oscillator proceeds with its natural frequency, while, when the coupling exceeds a certain threshold, a group of oscillators synchronize.

This phenomenon amazed Kuramoto, prompting him to begin working on collective synchronization phenomena with his first paper in 1975~\cite{Kuramoto75}.

In his next paper on the subject Kuramoto showed that, for any system of nearly identical, weakly coupled limit-cycle oscillators, the long-time dynamics is given by phase equations of the type
\eq{
\dot{\theta_i} =\omega_i + \sum_{j=1}^N \Gamma_{ij} (\theta_j - \theta_i), \hspace{1cm} i = 1, ..., N.
}
where we have indicated with $\theta_i$ the phase of the $i$-th oscillator, with $\omega_i$ its natural frequency, while the interaction function is $\Gamma_{ij}$~\cite{Kuramoto84}.

Although the model is drastically simplified, it remains quite difficult to analyze in general. Like Winfree, Kuramoto also recognized that the mean-field approximation was a way to arrive at a more tractable model.  He chose the coupling equally weighted and purely sinusoidal, that is,
\eq[Kuramoto]{
\Gamma_{ij} (\theta_j - \theta_i) = \frac{K}{N} \sin (\theta_j - \theta_i),
}
where $K$ represents the strength of the coupling and is obviously always positive ($K \ge 0$), while the factor $N$ in the denominator ensures that the model behaves well in the limit $N \rightarrow \infty$.

\begin{figure}[ht]
    \centering   \includegraphics[width=0.6\textwidth]{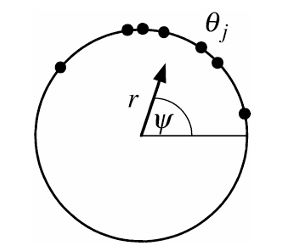}
    \caption{The visualization of a 7-particle Kuramoto system. The quantity $r$ indicates the distance of the center of mass from the center of the circle, and the angle $\psi$ denotes the relative phase.}
    \label{fig:cerchio}
\end{figure}

Through some additional assumptions and some algebraic steps it is possible to write the following equation
\eq{
\dot{\theta_i} =\omega_i +K r \sin (\psi - \theta_i),  \hspace{1cm} i = 1, ..., N.
}
where $r=r(t)$ is the radial distance (from the origin)  of the center of mass of rotators (assuming that the radius of the circle is unitary),
\eq{
   r^2  = \left( \sum_i \sin(\theta_i)\right)^2 + \left( \sum_i \cos(\theta_i)\right)^2,
}
and $\psi=\psi (t)$ is the corresponding phase, as shown in fig.~\ref{fig:cerchio}~\cite{Strogatz2000}.

Its interpretation could be as follows: if all the points are very close to each other, the radius is approximately 1 and the population behaves like a large oscillator; if, on the contrary, the points are distributed along the circumference, the radius is almost zero and the individual oscillations add up incoherently, failing to produce a macroscopic rhythm.

In this form, it appears that each oscillator is decoupled from the others, although in reality they are interacting through the quantities $r$ and $\psi$. When the coupling term is large, synchronization will occur; when it is small, it can be neglected with respect to $\omega_i$, and each oscillator will proceed at its natural frequency.

\subsection{Fireflies synchronization: educational proposal}

The aim of the game is to help students visualize how phase transitions work using Kuramoto's model applied to a context as familiar as possible, that of fireflies.

The idea that inspired us is the model \textit{Fireflies}, available on the Netlogo~\cite{NetLogoFireflies} platform website.

The model (\textit{Lucciole}) simulates the collective behavior of fireflies, implementing the Kuramoto model: each firefly changes its luminosity in a sinusoidal way, with a random natural frequency, but this is varied by a global coupling, as in \Eq{Kuramoto}. 

At the start of the game, the playground is populated with 100 fireflies arranged in a square grid, each of which turns on and off at its own frequency.

Each firefly can communicate with the others, and their effectiveness in doing so is given by the  parameter $K$, which can be varied via a slider.

The player's goal is to understand the relationship between communication capacity (the parameter $K$) and the firefly synchronization, which can be observed through three distinct channels:
\begin{itemize}
    \item Visual inspection. Students can observe the fireflies flashing and thus intuitively understand when they synchronize.
    \item Plot. By observing the graph where the  synchronization indicators (the $r$ of the Kuramoto model) is plotted versus time, it is possible to understand when synchronization is reached since the function tends to a horizontal asymptote if $K \ge K_c$, otherwise, when $K < K_c$, an irregular trend is observed.
    \item Numerical value. The synchronization indicator $r$ is also shown numerically.
\end{itemize}  

Using these three different channels allows for a more "holistic" approach to the problem. The first channel, the one involving map observation, allows even those without sufficient physical-mathematical training to grasp what is happening. The other two allow for more in-depth scientific analysis.

The numeric channel allows, once a value has been chosen for which the fireflies are synchronized, to show that the value is  changing and that it is sufficient to increase the number of digits read to verify this. However, once the number of significant digits has been fixed, at a certain point a constant value will be displayed.

Finally, the channel that refers to the plot allows students to analyze the study of functions in a more robust, but also playful way, examining in particular the achievement of a limit value given by the asymptote.

\begin{figure}[ht]
    \centering
    \includegraphics[width=0.8\textwidth]{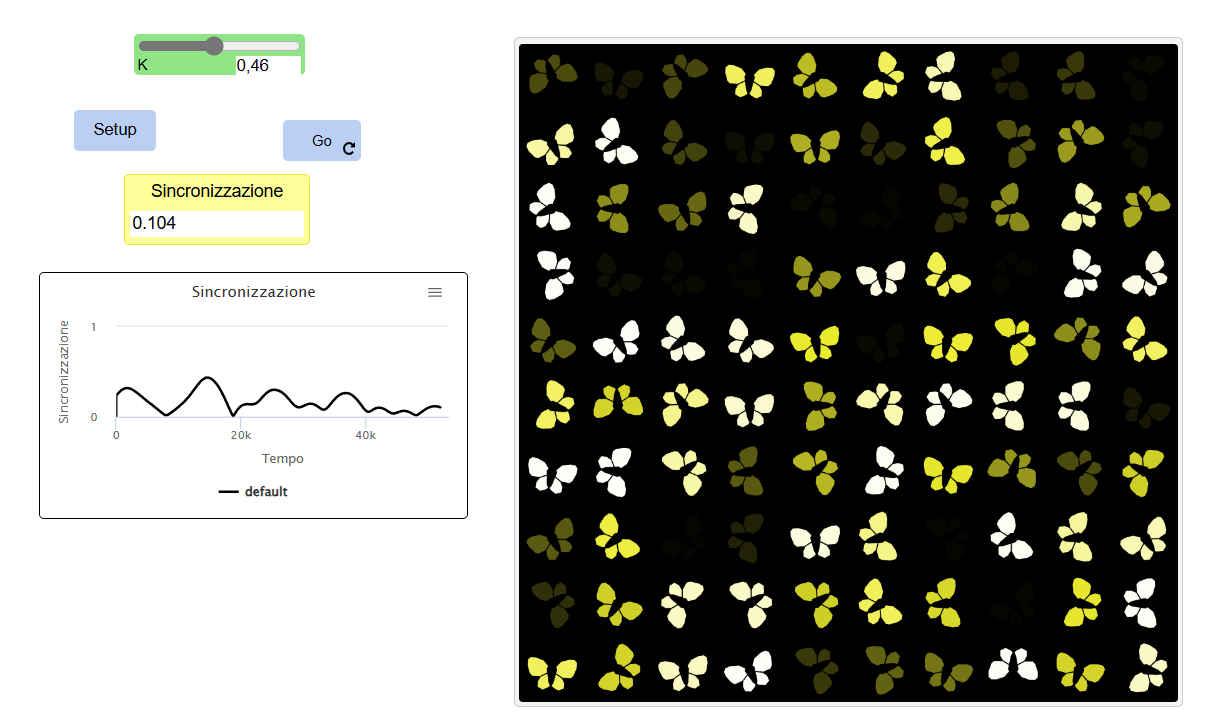}
    \caption{The firefly playground with $K$ smaller than the critical value $K_c$. One can see visually, using the plot and the synchronization parameter $r$ that the fireflies are not flashing synchronously.}
    \label{fig:L1}
\end{figure}

\begin{figure}[ht]
    \centering
    \includegraphics[width=0.8\textwidth]{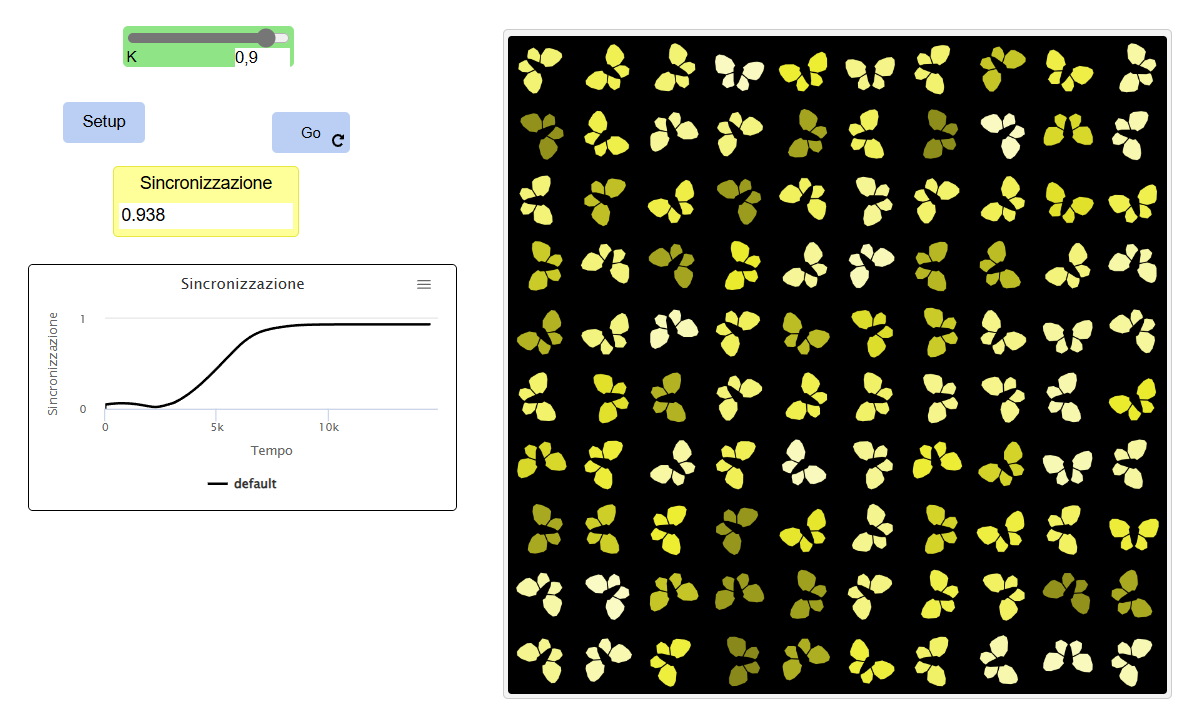}
    \caption{The firefly playground with $K$ larger than the critical value $K_c$. One can see visually, using the plot and the synchronization parameter $r$ that the fireflies are now synchronized.}
    \label{fig:L2}
\end{figure}

The test of the game was the request to estimate the threshold value $K_c$ for which the fireflies are synchronized and can therefore turn on and off simultaneously. To do this, it was suggested to observe what happens for small and large values of $K$ (as shown in Fig. ~\ref{fig:L1} and \ref{fig:L2}), thus leading them to the intuition that there must be a value for which this transition occurs.

\subsection{In-dept considerations for the Fireflies synchronization}\label{sec:fireflies-depth}

The discussion about the consequences of synchronization can start from the vision of Strogatz's TED talk~\cite{synchronization}, which presents examples of this phenomenon in nature, an experimental realization with metronomes, and the analysis of consequences on the millennium bridge. 

The teacher should put in evidence that this phenomenon concerns \textit{non-linear} oscillators, or oscillators coupled in non-linear ways. If one linearly couples two pendulums (that for small amplitudes behave as harmonic oscillators), one gets beats~\cite{coupled}, as when tuning a chord instrument. The reason is that for such a system there are two normal modes, one corresponding to them going together and the other in opposition, with slightly different frequencies. A generic initial condition is a superposition of them. But if one for instance links them with an inextensible string, so that their opposite motion is forbidden, the synchronous motion is the only remaining.

\section{Results}\label{sec:results}

We collected the numerical answers to test games adapting  a  NetLogo extension, as reported in Sect.~\ref{sec:NetLogo}, and these answers served as a starting point for subsequent discussions. 

We also conducted qualitative analysis of the responses received from students. Responses were provided in two ways (both anonymously): the first through open-ended questions on paper, asked directly during the administration of the games; the second through a Google form, although its compilation was volunteer.

It is obvious that the second method provided us with a significantly smaller number of responses, but with the advantage of a much faster analysis of the responses themselves.

This difficulty was partially resolved by asking some questions through both methods to be more certain of being able to receive them.

The analysis we will propose therefore combines the two types of responses, sometimes not fully consistent with each other.

The administration of the test and the collection of (anonymous) responses have been carried out by teachers, as part of their teaching program, so the ethical committee approval is not needed.

We managed to propose the games to all classes, from first to fifth, a Florence high school comprising both a scientific curriculum and one specialized in humanities, with an economic and social option, obtaining 113 responses through the Google form and 158 responses through the paper form.

The school made two school hours available (each school hour is 55 minutes) for each class and the games were played in the school's computer room.

Once the class was moved to the computer room, where the computers were usually already turned on, the teacher presented a brief explanation of the activity that would be carried out, also informing them that they would be able to answer some questions anonymously.

To ensure that responses were provided without the use of sensitive data, each student was randomly assigned a pre-prepared alphanumeric code. This code was then used to identify them both on paper and via  Google form.

After  their IDs were assigned by the teacher,  the activity could start. Before each game, each class member was given sheets of paper containing  questions related to each game.

Students were also given a sheet of paper where they could write down some answers to questions that would then be submitted via a Google form. Unfortunately, this idea only occurred to us after the first activities, after we noticed that students didn't quite remember the answers unless written on the sheet.

The following games were played in chronological order:
\begin{itemize}
    \item \textit{A site percolation} game titled \textit{forest fire};
    \item \textit{A mixed percolation} game titled \textit{forest fire with probability};
    \item \textit{A percolation test game} titled \textit{forest fire test};
    \item \textit{A sensitivity dependence to initial conditions} game titled \textit{chaotic pendulum};
    \item \textit{A test} related to the previous game titled \textit{chaotic pendulum test};
    \item \textit{A game/test about the Kuramoto model} titled \textit{fireflies synchronization}.
\end{itemize}

Before each game, the rules and instructions were listed, leaving room for questions. Whenever possible, teachers tried to engage the class, after the game, by asking questions or pointing out specific aspects, such as possible similarities between percolation models and epidemiological models, or by looking together for examples of chaotic physical systems.

At the end of the activity, the Google form was made available, kindly asking to complete it as soon as possible.

One of the biggest difficulties encountered in the implementation of the project is certainly due to the time factor: the two school hours, which correspond to one hour and fifty minutes, become in fact less since time was somehow lost by the class in getting to the computer room, sitting down, as well as in preparing for the next lesson.

Part of the time was devoted to questions, in our case in written form, so they could be collected. However, within an already underway project, the same questions can be asked orally to create a climate of active discussion.

To carry out a \textit{gaming} activity of this type, we estimate that at least the double of the number of hours are necessary, in order to give students the opportunity to become familiar with the games, create a stronger bond between teacher and class and give more space to discussion and \textit{peer instruction}.

Another critical issue was the space factor, given that the computer room could barely hold a class, and the computer stations were very close  each other.

This, in addition to the fact that it was one of the rare experiences of computer room activities for them, often led to lapses in concentration.

In hindsight, it probably would have been more interesting to conduct a survey considering smaller numbers of players, giving them more time.

\subsection{Qualitative analysis of the responses received through the Google form}

\begin{figure}[ht]
    \centering
    \includegraphics[width=0.8\textwidth]{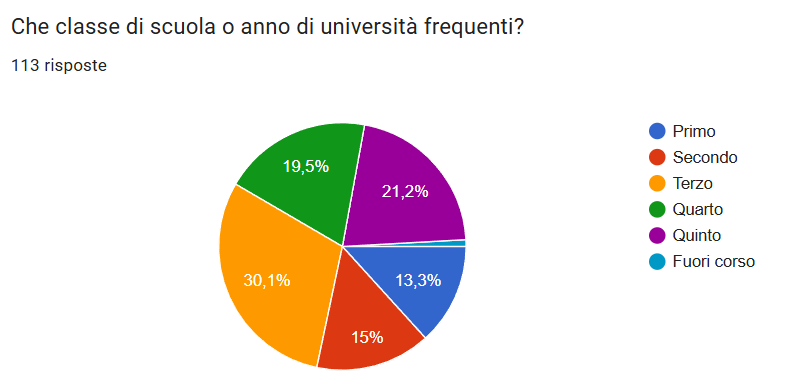}
    \caption{The gender breakdown (percentage of females) based on school years. It results to be quite homogeneous.}
    \label{fig:Scuola}
\end{figure}

Among the people who responded to the Google form, 60 declared themselves to be male, 48 declared themselves to be female, 3 declared themselves to be non-binary, while 2 people did not want to declare it, as shown in Table~\ref{tab:Tabellagenere}. Instead, in fig.~\ref{fig:Scuola} is shown the gender breakdown by school year and curricula.

\begin{table}[ht]
    \centering
    \includegraphics[width=0.8\textwidth]{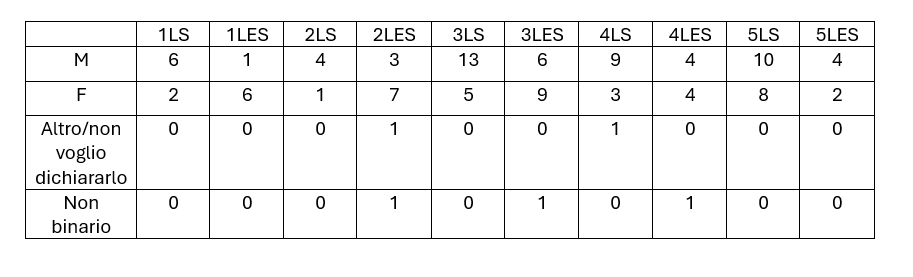}
    \caption{Gender breakdown by school year and curricula. LS stands for scientific high school, while LES stands for economic and social high school.}
    \label{tab:Tabellagenere}
\end{table}

Among the questions we asked via the online form were some specific ones about the use of videogames or board/role-playing games.

\begin{figure}[ht]
    \centering
    \begin{tabular}{cc}
    (a) & (b) \\
    \includegraphics[width=0.4\textwidth]{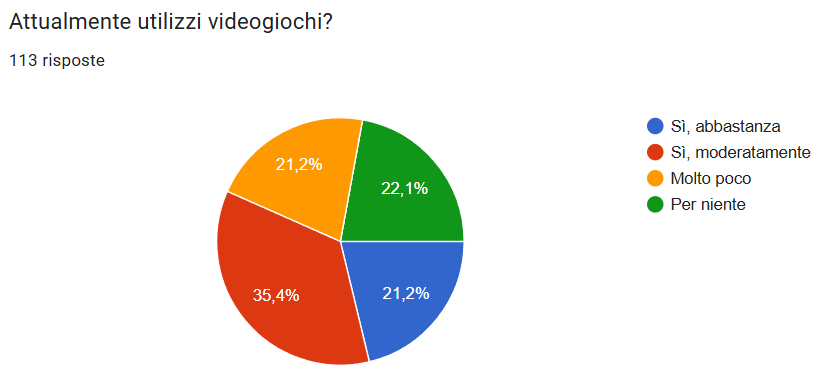} & 
    \includegraphics[width=0.4\textwidth]{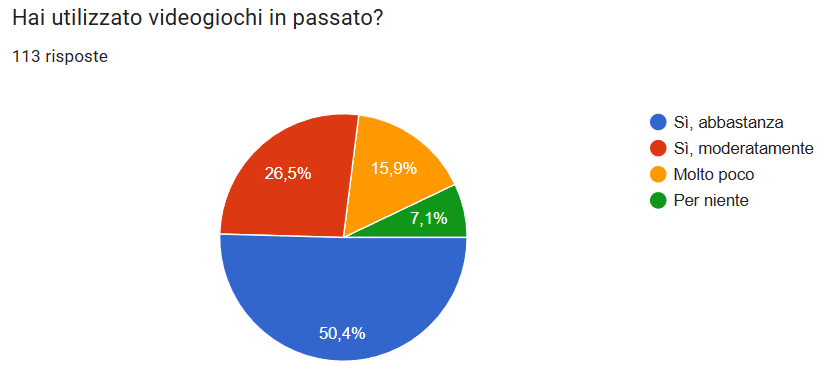}
    \end{tabular}
    \caption{Responses about the practice of videogames. (a) Present. (b) In the past. }
    \label{fig:VG}
\end{figure}

\begin{figure}[ht]
    \centering
    \begin{tabular}{cc}
    (a) & (b) \\
    \includegraphics[width=0.4\textwidth]{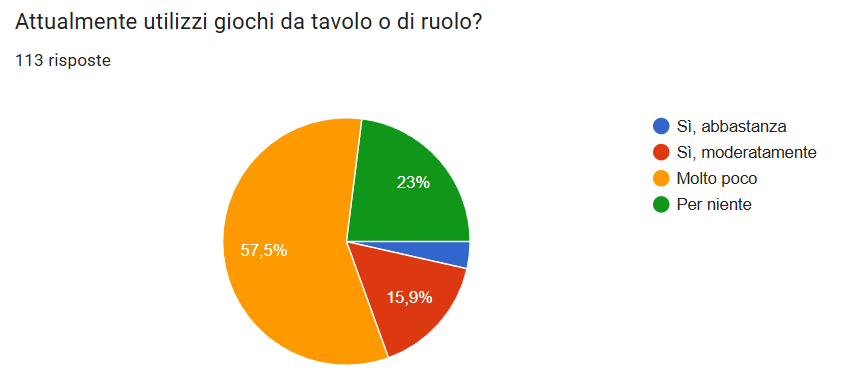} & 
    \includegraphics[width=0.4\textwidth]{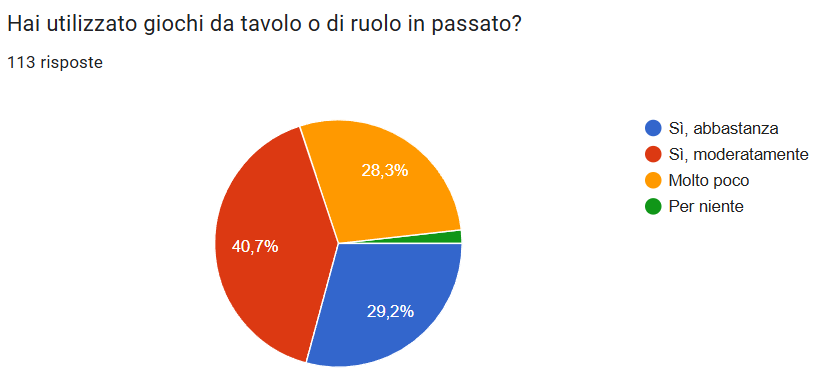}
    \end{tabular}
    \caption{Responses about the practice of board or role-playing games. (a) Present. (b) In the past. }
    \label{fig:GDT}
\end{figure}

For both types of games, a decrease in play activity is observed as male and female students grow older. Interestingly, videogames are still used more frequently than board games or role-playing games. For a clearer picture, see Figs.~\ref{fig:VG}, and \ref{fig:GDT}.

A section of the Google form was dedicated to surveying both the satisfaction with the proposed games and the difficulty encountered.

\begin{figure}[ht]
    \centering
    \begin{tabular}{cc}
    (a) & (b) \\
    \includegraphics[width=0.4\textwidth]{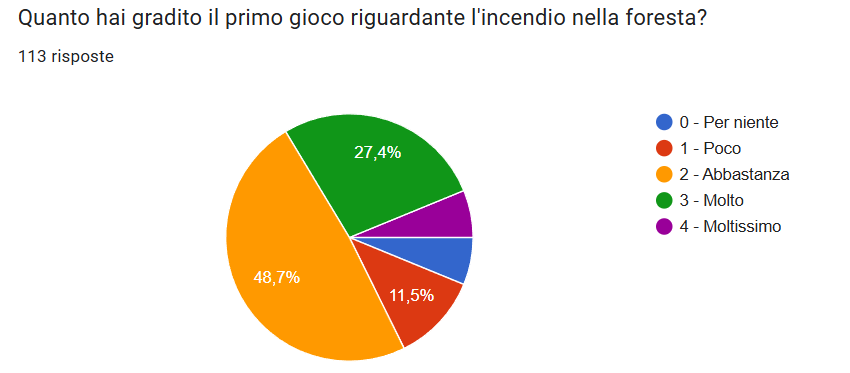} &
     \includegraphics[width=0.4\textwidth]{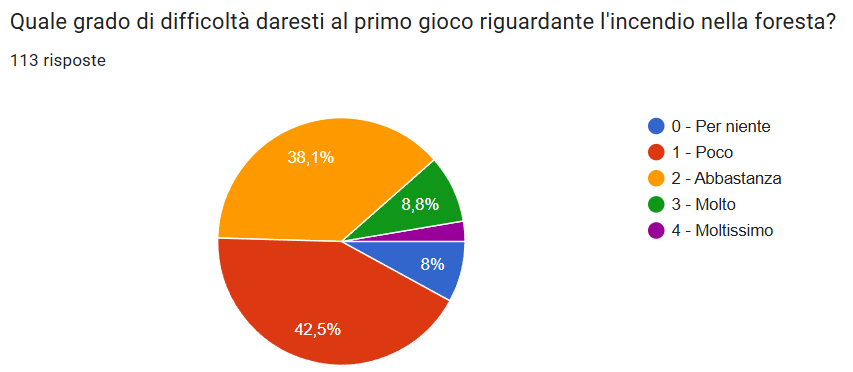}
     \end{tabular}
    \caption{Responses about the f\textit{orest fire} game.(a) Enjoyment. (b) Difficulties. }
    \label{fig:ForestFire}
\end{figure}

We can see that more than $80\%$ liked \textit{forest fire} quite a bit, enough, or a lot (see fig.~\ref{fig:ForestFire}-(a)), even though half of the players still found it difficult, as shown in fig.~\ref{fig:ForestFire}-(b).

\begin{figure}[ht]
    \centering
    \begin{tabular}{cc}
    (a) & (b) \\
    \includegraphics[width=0.4\textwidth]{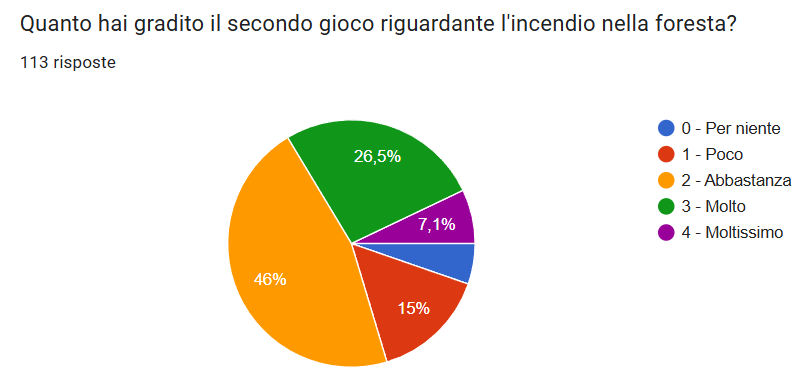} &
    \includegraphics[width=0.4\textwidth]{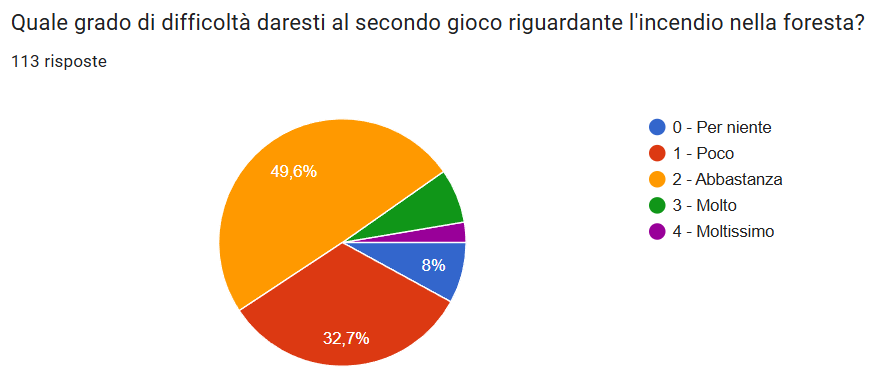}
    \end{tabular}
    \caption{Responses about the \textit{forest fire with probability} game. (a) Enjoyment. (b) Difficulties. }
    \label{fig:ForestFireProbability}
\end{figure}

The same results for the game \textit{forest fire with probability}, which has a similar approval rating, as shown in fig.~\ref{fig:ForestFireProbability}-(a) despite $60\%$ rating it as difficult (see Fig.~\ref{fig:ForestFireProbability}-(b)).

\begin{figure}[ht]
    \centering
    \begin{tabular}{cc}
    (a) & (b) \\
    \includegraphics[width=0.4\textwidth]{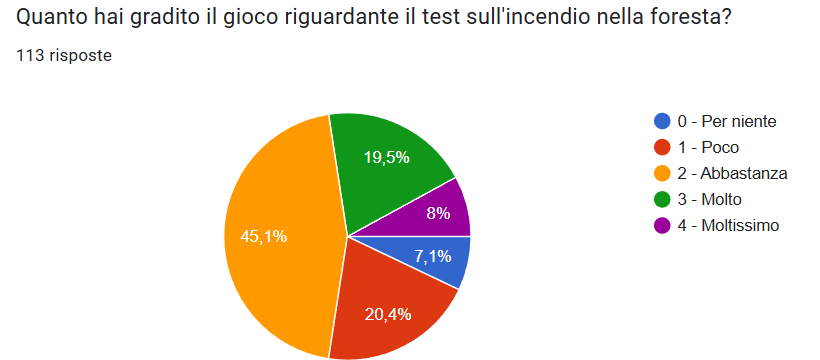}&
    \includegraphics[width=0.4\textwidth]{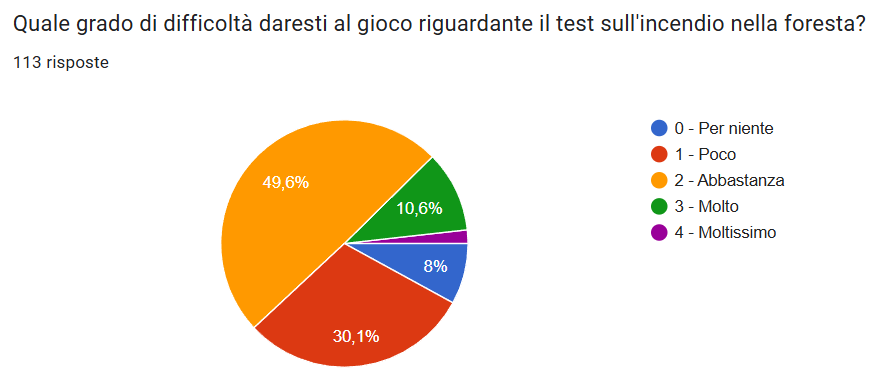}
    \end{tabular}
    \caption{Responses about  the \textit{forest fire} test game. (a) Enjoyment. (b) Difficulties. }
    \label{fig:ForestFireTest}
\end{figure}

We conclude with the \textit{forest fire} test game. This game also follows the same evaluation as the previous two in terms of both enjoyment and perceived difficulty, as one can see in Figs.~\ref{fig:ForestFireTest} .

\begin{figure}[ht]
    \centering
    \begin{tabular}{cc}
    (a) & (b) \\
    \includegraphics[width=0.4\textwidth]{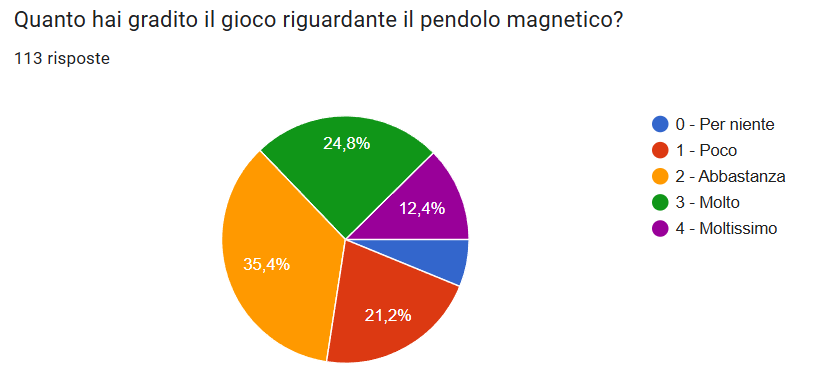} &
    \includegraphics[width=0.4\textwidth]{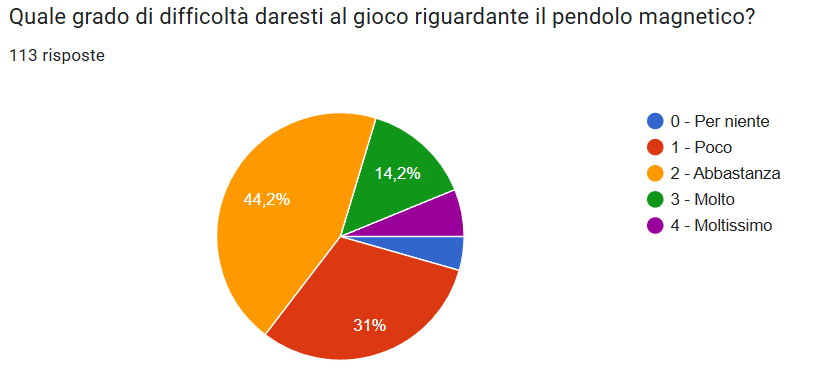}
    \end{tabular}
    \caption{Responses about the \textit{chaotic pendulum} game. (a) Enjoyment. (b) Difficulties. }
    \label{fig:pendulumGame}
\end{figure}

\begin{figure}[ht]
    \centering
    \begin{tabular}{cc}
    (a) & (b) \\
    \includegraphics[width=0.4\textwidth]{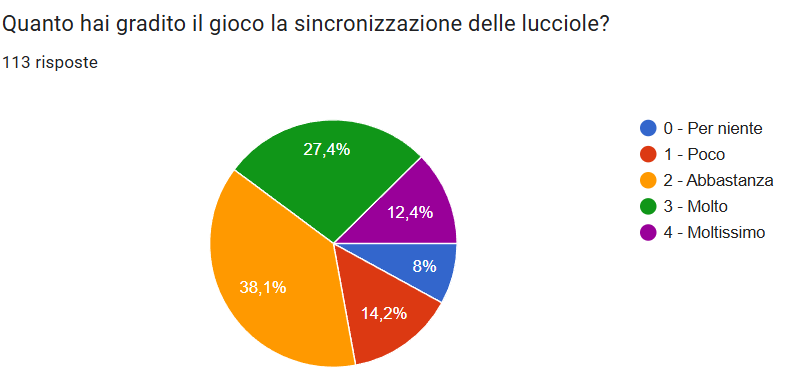}&
    \includegraphics[width=0.4\textwidth]{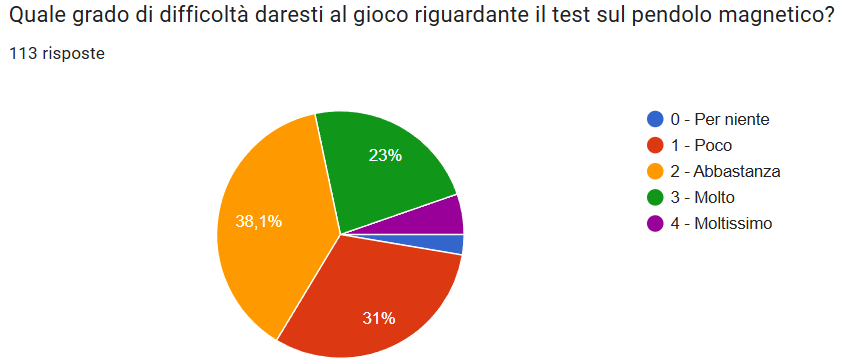}
    \end{tabular}
    \caption{Responses about the \textit{chaotic pendulum test} game. (a) Enjoyment. (b) Difficulties. }
    \label{fig:pendulumGameTest}
\end{figure}

Also the games related to sensitivity to initial conditions, i.e., the \textit{chaotic pendulum} had a positive response of more than $70\%$ regarding enjoyment as shown in Fig.~\ref{fig:pendulumGame}-(a), with similar percentages regarding perceived difficulty (see Figs.~\ref{fig:pendulumGame}-(b) and \ref{fig:pendulumGameTest}).

\begin{figure}[ht]
    \centering
    \begin{tabular}{cc}
    (a) & (b) \\
    \includegraphics[width=0.4\textwidth]{Gradimento5.png} & 
    \includegraphics[width=0.4\textwidth]{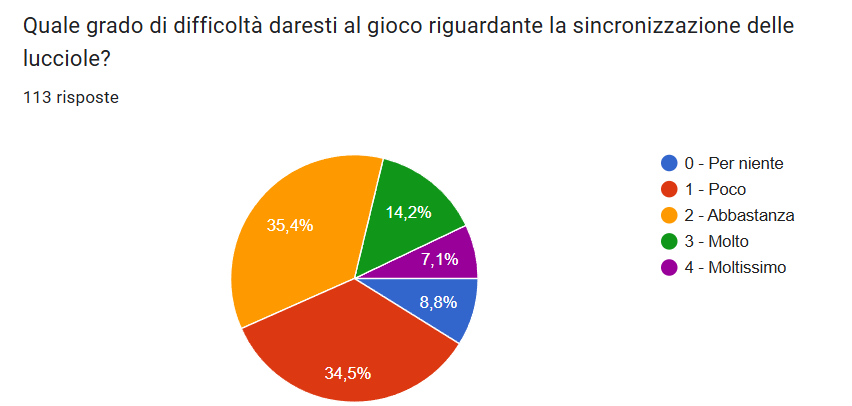}
    \end{tabular}
    \caption{Responses about the \textit{firefly synchronization} game. (a) Enjoyment. (b) Difficulties. }
    \label{fig:fireflyGame}
\end{figure}

The last game proposed, the one about fireflies synchronization, was liked quite a lot, much, or very much by almost $80\%$ of the players, as shown in fig.~\ref{fig:fireflyGame}-(a), although it was also difficult for more than half of the participants (see fig.~\ref{fig:fireflyGame}-(b)).

Analyzing the answers students gave to the questions asked, we can say that as far as the games related to \textit{forest fire} were concerned, almost all of them understood the relationship between the percentage of burned trees and the density (and also size) of the trees.

Many students also noted that there can be a high level of variability when trying to run the game using the same parameters, and that this variability is due to both the arrangement of the trees and the tree that starts the fire.

Below are some responses we received from players regarding what was just stated:
\begin{itemize}
    \item \textit{In some cases, the percentage of burned trees remains the same; in other cases, changing the arrangement changes the value. The higher the density, the more trees will burn, and the percentage remains the same. The lower the density, the fewer trees will burn, and the percentage will not change. If we stick to a medium density, the values will change more.}
    \item \textit{It can be noted that, despite the same tree density value, depending on the point where the fire started, there will be a  low percentage of burned trees.}
    \item \textit{Density doesn't necessarily lead to a higher result, but only to potential damage. The same value produces different results precisely because of a different arrangement between one test and the next.}
\end{itemize}

A small portion of the responses revealed difficulties in understanding what a relationship between quantities is. For example, some students stated that the percentage of burned trees increased if their density increased, subsequently responding that there was no relationship between the two quantities, thus giving rise to a contradiction.

One explanation may be that some of them probably only know a few types of mathematical relationships; in particular they often refer only to direct or inverse proportional relationships.

Although we didn't dwell on the possible relationship between fire expansion and its propagation time during the experiment, some students emphasized this very connection. Below are some of their responses:
\begin{itemize}
    \item \textit{In my opinion, this game is useful to understand how long it would take and how many trees would burn so as to predict it in real life.}
    \item \textit{We can understand that the placement of trees is essential to reduce the percentage of burned trees and that acting in the first minutes of a fire can prevent a catastrophe.}
    \item \textit{It is used to calculate how long it takes for a fire to spread through a forest.}
\end{itemize}

When asking how the forest fire games worked and what they meant, some interesting answers emerged, which we decided to report:
\begin{itemize}
    \item \textit{Its function is to demonstrate how a small variation can lead to different results.}
    \item \textit{This game somewhat reflects today's civilization in that with a huge number of people as soon as something happens it spreads very quickly, from fashion to diseases.}
    \item \textit{Controlled planting or clearing can prevent/control a fire.}
\end{itemize}

We were also pleased to read a comment from a student, who stated that the purpose of the game is \textit{to explain physics by finding an alternative way} and \textit{to learn physics better}.

Interestingly, in the forest fire and magnetic pendulum games/tests, some students claim to choose a radius value or a position "randomly," but we suspect that this response is due to their poor ability to express themselves: they are not actually making a random choice but, while reasoning about their decision, they do not know how to express the logical-deductive process that led them to that choice.

\begin{figure}
    \centering
    \includegraphics[width=0.5\linewidth]{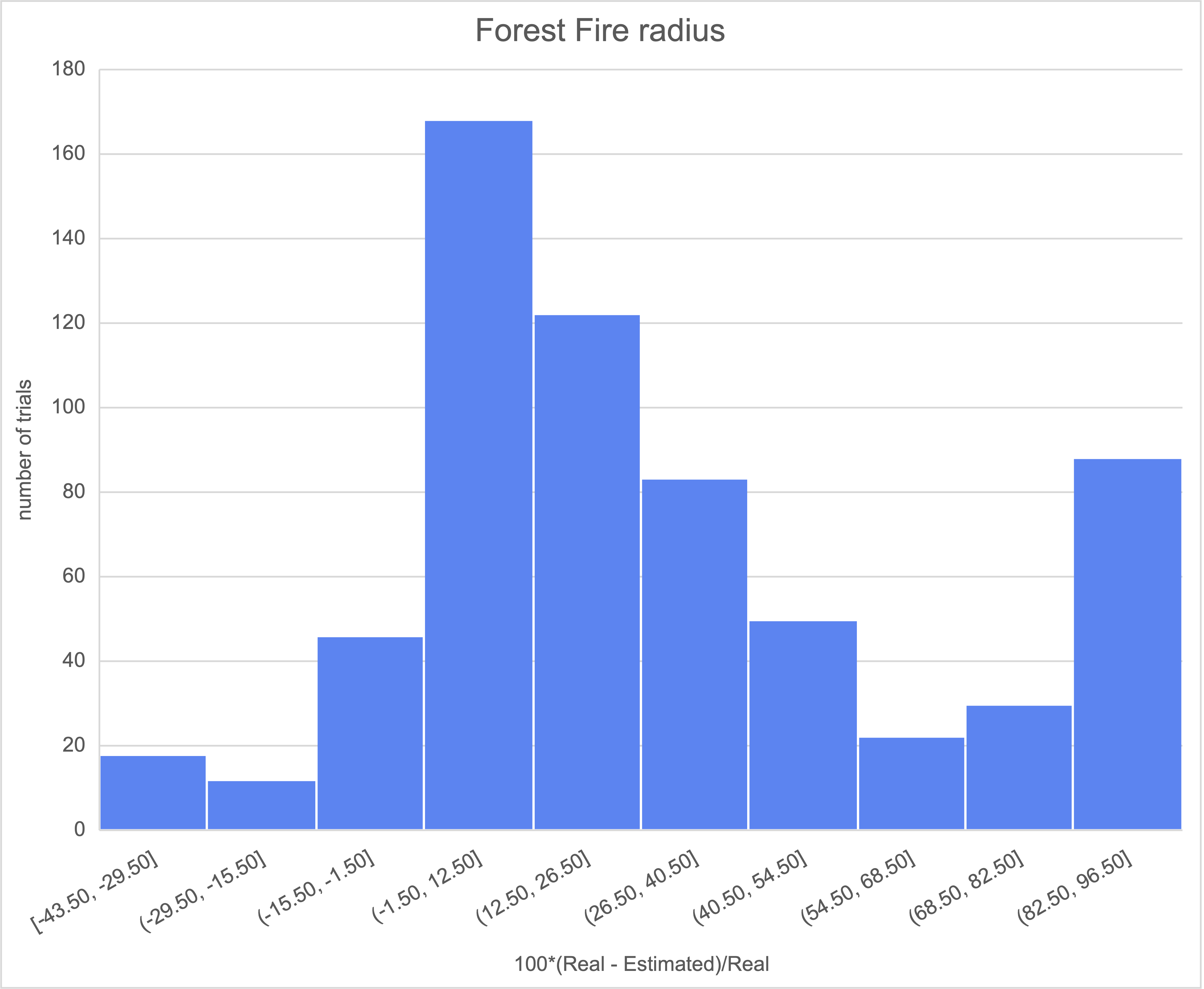}
    \caption{Numerical results from the \textit{forest fire with probability} game. The histogram shows the number of trials (in average $5.5$ for each student) corresponding to the percentage of the distance from the "real" value. While there is a peak at a reasonable intermediate value, there are also peaks at the extreme values. }
    \label{fig:ForestFireTest1}
\end{figure}

It is also possible that they did not understood how to manipulate the interface. Indeed, from Fig.~\ref{fig:ForestFireTest1}, one can see that there are ``extreme'' peaks in the distribution of chosen radii (corresponding essentially to a minimum or maximum values allowed by the game), indicating a difficulty in playing the game. 

Among those who justify their choice, noteworthy responses emerge, such as the following:
\begin{itemize}
    \item \textit{I placed the circumference where the density was decreasing so it was more difficult for the fire to spread even further.}
    \item \textit{I chose a radius of 19.6 because it captured an area where the trees were still close together.}
    \item \textit{Since the burnt tree in the center was surrounded by large, close-knit trees, they would certainly have burned, while the ones further away were small and far apart, so it wouldn't have spread as much.}
\end{itemize}

Even regarding the magnetic pendulum game, the responses suggest that almost everyone has understood its fundamental characteristic: the sensitivity to the initial conditions of some physical systems. For example, when asked what they noticed from simulations in which the pendulum starts from slightly different locations, many responded that \textit{in some cases it returns to the same magnet, but sometimes this varies. This doesn't happen if the starting point is too close to a magnet.}
Others claim that the behavior is due to the fact that \textit{there are strange magnetic fields} or that \textit{physics is unpredictable}.

It is then interesting to analyze the choices made by students in testing this game: many correctly responded that the areas which are most sensitive to the initial conditions are either near the  the center of the map, or peripheral areas, or on the line segment equidistant between two magnets.

\begin{figure}
    \centering
    \includegraphics[width=0.4\linewidth]{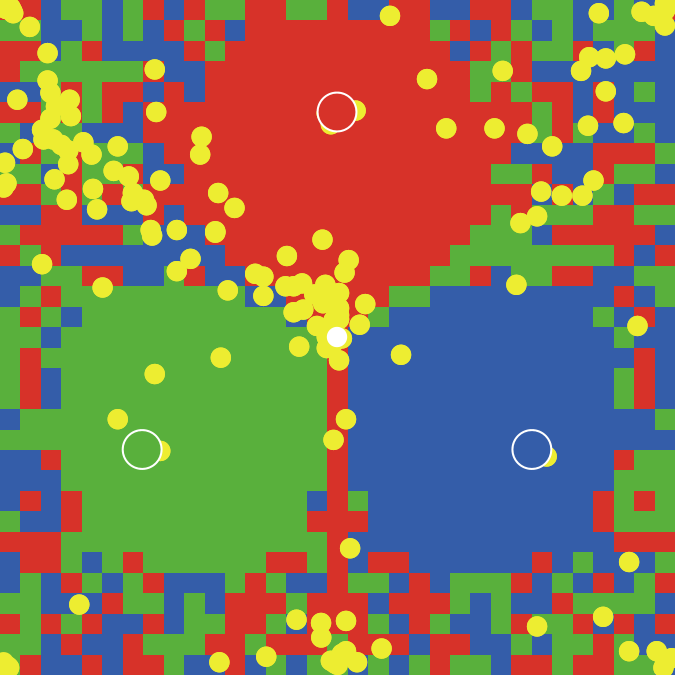}
    \caption{Distribution of the points reputed more sensible, alongside the basin of attraction of the chaotic pendulum game.}
    \label{fig:distribution}
\end{figure}

One can see in Fig.~\ref{fig:distribution}  the distribution of the points on the map reputed more sensitive, alongside the basins of attraction in the background. There is an evident preference for the top-right portion of the graph, probably related to visual scanning pattern left-right and top-bottom typical of reading. 

\begin{figure}
    \centering
    \begin{tabular}{cc}
    (a) & (b) \\
    \includegraphics[width=0.4\linewidth]{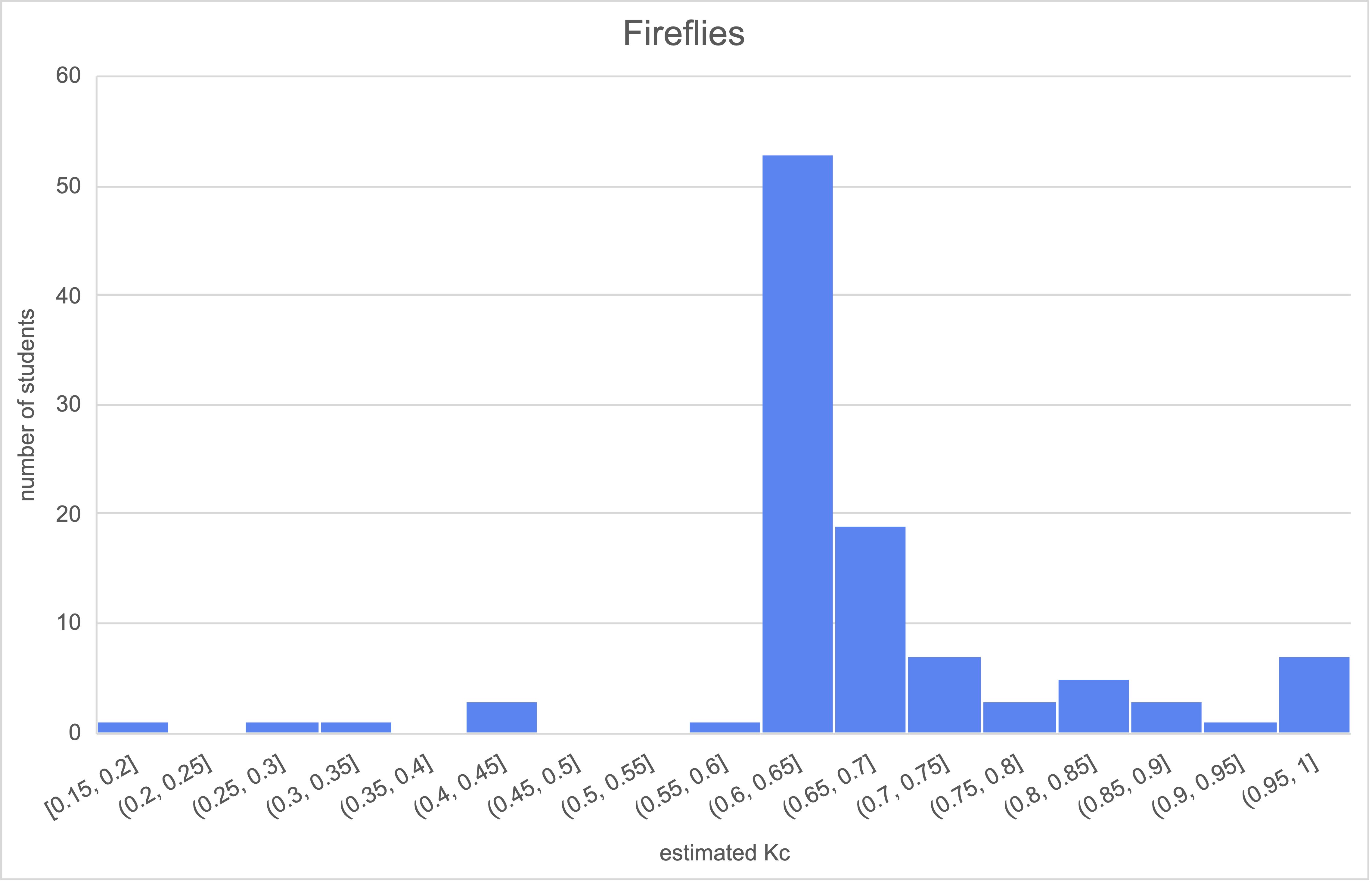} &
    \includegraphics[width=0.4\linewidth]{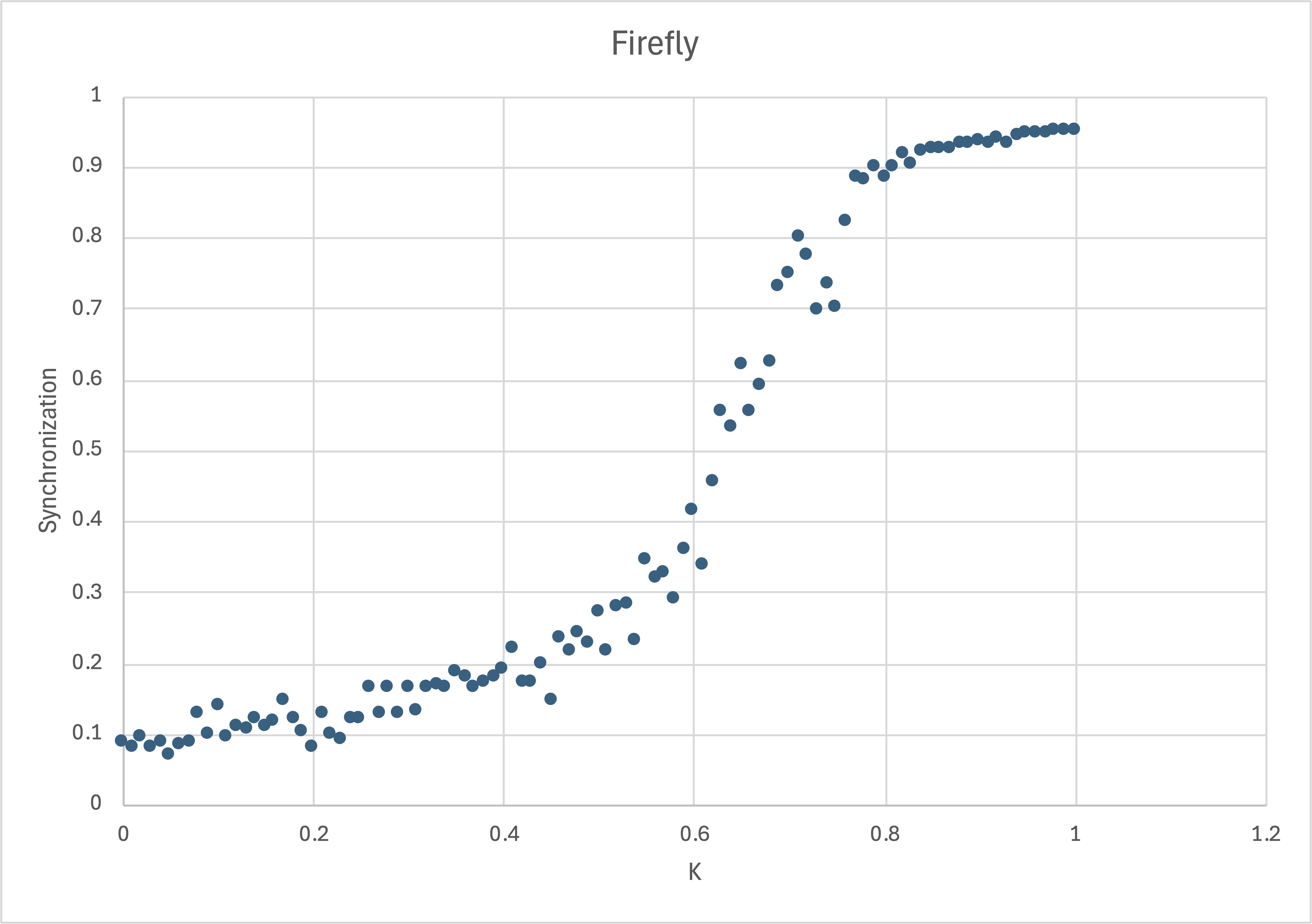}
    \end{tabular}
    \caption{Results of the firefly game test. (a) The synchronization indicator $r$ as a function of the synchronization parameter $K$ (average over 10 runs). (b) The histogram of the students' choice.}
    \label{fig:fireflies-test}
\end{figure}

As shown in Fig.~\ref{fig:fireflies-test}, the distribution of the $K_c$ estimate done by 106 students (with additional 6 "not remembering" null answers) is peaked around reasonable values ($0.6\le K_c\le 0.75$), indicating that most of students understood the game.  

We also prepared  a second test game, with an additional parameter (varying radius of communication among fireflies), but we then realized that it would have needed too much time. So we collected the values of the estimated $K_c$, and therefore this single game served both as the introduction and the test.

Some of the students demonstrate that they have understood the transition phenomenon that leads the fireflies to synchronize, for example a student responds that \textit{depending on the given parameter $K$, the synchronization of the fireflies varies, and for $K<0.64$ the synchronization is not stable but decreases and therefore the fireflies are not synchronized; while for $K>0.64$ I realized that the fireflies will  gradually synchronize}.

Some also noted that the higher the value of the $K$ parameter (larger than the critical value), the more synchronized the fireflies are.

Other students declared that in their opinion the aim of the game is to \textit{"observe stability on a graph"} or that \textit{"it allows you to obtain, like an optimization problem, the minimum value for  the synchronization of the elements of the system"}.

\section{NetLogo implementation}\label{sec:NetLogo}

After that their IDs were assigned by the teacher,  the activity could start. Before each game, each class member was given sheets of paper containing  questions related to each game, for collecting opinions and answers, for those not willing to use the automatic recording described below.

We used the NetLogo~\cite{NetLogo} platform for all games, because it is easy to customize, self-contained (so that students could eventually modify it), and the model can also be exported in html + javascript~\cite{NetLogoWeb} so that it can be run on a browser (i.e., using a cellular phone or a tablet), without the need of downloading the application, even though some games (like the chaotic pendulum) do not play smoothly in such a case. 

We also developed a way of submitting the user id and  answers to questions directly from the application to a Google form. 

One first sets up the Google form, and uses the option to generate a pre-compiled form. In this way one gets an URL with a query containing the attributes identifying the fields of the form, i.e., something like
\\
\\
\noindent \verb|https://docs.google.com/forms/d/e/<FORMID>/viewform?|

\noindent \verb|usp=pp_url&entry.1526793516=one|
\\
\\
\noindent where \verb|<FORMID>| is a long string identifying the form, and \verb|entry.1526793516| identifies the field, recognized by the pre-compiled answer ``\verb|one|''. 

One can therefore programmatically obtain the values of the fields and build the query.
In order to submit the form simply visiting the URL, one has to replace \verb|viewform| with \verb|formResponse| and add \verb|&Submit=submit| at the end. 
In summary, assuming that the answer to first form is \verb|125|, one should build the query
\\
\\
\noindent \verb|https://docs.google.com/forms/d/e/<FORMID>/formResponse?|

\noindent \verb|usp=pp_url&entry.1526793516=125&Submit=submit|
\\

NetLogo provides the \verb|fetch| extension~\cite{fetch}, which was originally planned for dowloading material from the web, but it can also used to submit the compiled form. 

The html version of the app, however, shows an error message ("Extension exception") upon submitting the form, so we had to write a small shell script to remove such a message from the generated html files. 

\section{Conclusions}

The aim of this work was to investigate a teaching methodology different from the frontal one,  by applying a playful approach through the use of videogames.

To this end, we developed several videogames whose central scientific topics were not included in the secondary school curriculum, such as nonlinear phenomena, sensitivity to initial conditions, and phase transitions.

One reason we believe this type of teaching approach can be useful to students is that it encourages the recognition of common patterns in phenomena related to disciplines considered very different from each other, especially in a school context.

We also believe that the failure to address nonlinear phenomena, often present in everyday life, could create the impression that the physics taught and learned in school is completely disconnected from reality.

A possible foundation for this approach can be found in diSessa’s theory of conceptual ecology~\cite{Disessa2002,diSessa1998}. In this framework, students’ understanding is built from a network of intuitive knowledge elements shaped by everyday experience. As presented in this work, video-games that simulate nonlinear behaviors, sensitivity to initial conditions, and phase transitions can activate these intuitive resources and support their reorganization. Rather than replacing students’ everyday ideas, the playful and interactive environment helps refine and integrate them into more coherent scientific concepts, acting as a bridge between informal reasoning and formal physics~\cite{Gauthier2024,Velasco2021}.

Our work is intended to be an exploratory investigation, therefore the analysis carried out was not quantitative but only qualitative because we wanted to focus more on the feasibility of the proposed investigation, reserving the right to carry out a different treatment in the future.

From the analysis of satisfaction and difficulties encountered during the activity, we are proud to say that all the games proposed were enjoyed, despite the obstacles that students reported having encountered in playing them.

The reviews, both positive and negative, reveal both a desire to engage in school activities outside of classroom teaching and a desire to have fun while learning. On the other hand, we must admit that for some students, the activities were not very clear, and some would have preferred more explanations and less freedom of expression.

Both from the responses received and from our impressions, some doubts and critical issues regarding the work carried out emerge.

One of the biggest challenges was to squeeze the activity into less than two hours, and a structured activity of this kind could probably take twice as long to be carried out as optimally as possible.

More hours available could provide students with the opportunity to become more familiar with the dynamics of the game, also creating a stronger bond between teacher and class.

Similar games have also been proposed in non-formal education contexts such as at the Game festival ``Play Festival del Gioco 2025'' in Bologna  and "Lucca Comix and Games 2025" in Lucca.

In the future, we plan to continue offering the activity to other schools, requesting longer time slots or dividing the activity into multiple game sessions, and focusing more on the educational aspects (for example, leaving more space for discussion and peer instruction) and play, reducing the number of written questions.

We also plan to carry out a controlled study with pre- and post-testing to experimentally verify the effectiveness of this intervention.

\printbibliography

\end{document}